\documentclass[twoside,twocolumn,9pt]{article}
\usepackage{extsizes}
\usepackage[super,sort&compress,comma]{natbib} 
\usepackage[version=3]{mhchem}
\usepackage[left=1.5cm, right=1.5cm, top=1.785cm, bottom=2.0cm]{geometry}
\usepackage{balance}
\usepackage{mathptmx}
\usepackage{sectsty}
\usepackage{graphicx,multirow} 
\usepackage{lastpage}
\usepackage[format=plain,justification=justified,singlelinecheck=false,font={stretch=1.125,small,sf},labelfont=bf,labelsep=space]{caption}
\usepackage{float}
\usepackage{fancyhdr}
\usepackage{fnpos}
\usepackage[english]{babel}
\addto{\captionsenglish}{%
  
}
\usepackage{array}
\usepackage{droidsans}
\usepackage{charter}
\usepackage[T1]{fontenc}
\usepackage[usenames,dvipsnames]{xcolor}
\usepackage{setspace}
\usepackage[skipempty]{credits}
\usepackage[compact]{titlesec}
\usepackage{hyperref}

\definecolor{cream}{RGB}{222,217,201}

\begin{document}

\pagestyle{fancy}
\thispagestyle{plain}
\fancypagestyle{plain}{
\renewcommand{\headrulewidth}{0pt}
}

\makeFNbottom
\makeatletter
\renewcommand\LARGE{\@setfontsize\LARGE{15pt}{17}}
\renewcommand\Large{\@setfontsize\Large{12pt}{14}}
\renewcommand\large{\@setfontsize\large{10pt}{12}}
\renewcommand\footnotesize{\@setfontsize\footnotesize{7pt}{10}}
\makeatother

\renewcommand{\thefootnote}{\fnsymbol{footnote}}
\renewcommand\footnoterule{\vspace*{1pt}%
\color{cream}\hrule width 3.5in height 0.4pt \color{black}\vspace*{5pt}} 
\setcounter{secnumdepth}{5}

\makeatletter 
\renewcommand\@biblabel[1]{#1}            
\renewcommand\@makefntext[1]%
{\noindent\makebox[0pt][r]{\@thefnmark\,}#1}
\makeatother 
\renewcommand{\figurename}{\small{Fig.}~}
\sectionfont{\sffamily\Large}
\subsectionfont{\normalsize}
\subsubsectionfont{\bf}
\setstretch{1.125} 
\setlength{\skip\footins}{0.8cm}
\setlength{\footnotesep}{0.25cm}
\setlength{\jot}{10pt}
\titlespacing*{\section}{0pt}{4pt}{4pt}
\titlespacing*{\subsection}{0pt}{15pt}{1pt}

\fancyfoot{}
\fancyfoot[LO,RE]{\vspace{-7.1pt}\includegraphics[height=9pt]{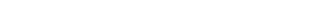}}
\fancyfoot[CO]{\vspace{-7.1pt}\hspace{13.2cm}\includegraphics{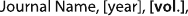}}
\fancyfoot[CE]{\vspace{-7.2pt}\hspace{-14.2cm}\includegraphics{head_foot/RF}}
\fancyfoot[RO]{\footnotesize{\sffamily{1--\pageref{LastPage} ~\textbar  \hspace{2pt}\thepage}}}
\fancyfoot[LE]{\footnotesize{\sffamily{\thepage~\textbar\hspace{3.45cm} 1--\pageref{LastPage}}}}
\fancyhead{}
\renewcommand{\headrulewidth}{0pt} 
\renewcommand{\footrulewidth}{0pt}
\setlength{\arrayrulewidth}{1pt}
\setlength{\columnsep}{6.5mm}
\setlength\bibsep{1pt}

\makeatletter 
\newlength{\figrulesep} 
\setlength{\figrulesep}{0.5\textfloatsep} 

\newcommand{\topfigrule}{\vspace*{-1pt}%
\noindent{\color{cream}\rule[-\figrulesep]{\columnwidth}{1.5pt}} }

\newcommand{\botfigrule}{\vspace*{-2pt}%
\noindent{\color{cream}\rule[\figrulesep]{\columnwidth}{1.5pt}} }

\newcommand{\dblfigrule}{\vspace*{-1pt}%
\noindent{\color{cream}\rule[-\figrulesep]{\textwidth}{1.5pt}} }

\makeatother

\twocolumn[
  \begin{@twocolumnfalse}
{\includegraphics[height=30pt]{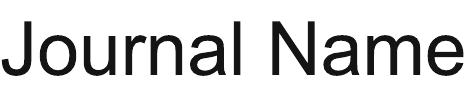}\hfill\raisebox{0pt}[0pt][0pt]{\includegraphics[height=55pt]{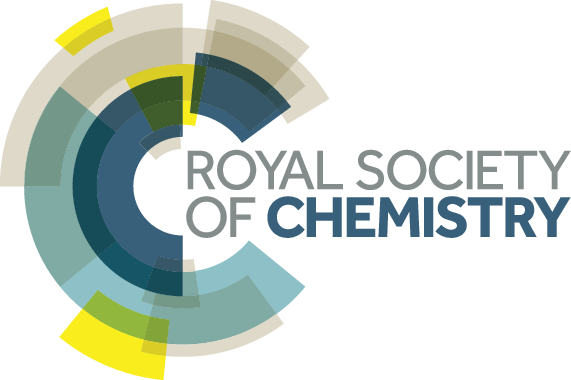}}\\[1ex]
\includegraphics[width=18.5cm]{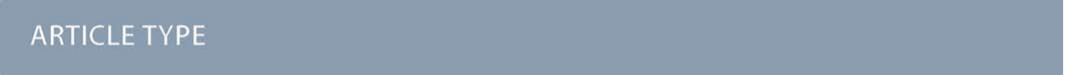}}\par
\vspace{1em}
\sffamily
\begin{tabular}{m{4.5cm} p{13.5cm} }

\includegraphics{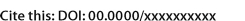} & \noindent\LARGE{\textbf{Flux and fluence effects on the Vacuum-UV photodesorption and photoprocessing of CO$_2$ ices$^\dag$}} \\
\vspace{0.3cm} & \vspace{0.3cm} \\

 & \noindent\large{Antoine Hacquard,$^{\ast}${$^{ab}$} Daniela Torres-Diaz,{$^{ac}$} Romain  Basalgète,{$^{b \ddag}$} Delfina Toulouse,{$^{b}$}  Géraldine Féraud,{$^{a}$} Samuel Del Fré,{$^{d}$} Jennifer A. Noble,{$^{e}$} Laurent Philippe,{$^{a}$} Xavier Michaut,{$^{a}$} Jean-Hugues Fillion,{$^{a}$} Anne Lafosse,{$^{c}$} Lionel Amiaud,{$^{c}$} and Mathieu Bertin{$^{a}$}}\\

\includegraphics{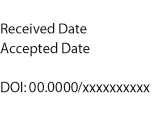} & \noindent\normalsize{CO$_2$ is a major component of the icy mantles surrounding dust grains in planet and star formation regions. Understanding its photodesorption is crucial for explaining gas phase abundances in the coldest environments of the interstellar medium irradiated by vacuum-UV (VUV) photons. Photodesorption yields determined experimentally from CO$_2$ samples grown at low temperatures (T=15~K) have been found to be very sensitive to experimental methods and conditions. Several mechanisms have been suggested for explaining the desorption of CO$_2$, O$_2$ and CO from CO$_2$ ices. 
In the present study, the cross sections characterizing the dynamics of photodesorption as a function of photon fluence (determined from released molecules in the gas phase) and of ice composition modification (determined \textit{in situ} in the solid phase) are compared for the first time for different photon flux conditions (from 7.3$\times 10^{12}$~photon/s/cm$^2$ to 2.2$\times 10^{14}$~photon/s/cm$^2$) using monochromatic synchrotron radiation in the VUV range (on the DESIRS beamline at SOLEIL).
This approach reveals that CO and O$_2$ desorption are decorrelated from that of CO$_2$. CO and O$_2$ photodesorption yields depend on photon flux conditions and can be linked to surface chemistry. By contrast, the phodesorption yield of CO$_2$ is independent of the photon flux conditions and can be linked to bulk ice chemical modification, consistently with an indirect desorption induced by electronic transition (DIET) process.}

\end{tabular}

 \end{@twocolumnfalse} \vspace{0.6cm}
  ]

\renewcommand*\rmdefault{bch}\normalfont\upshape
\rmfamily
\section*{}
\vspace{-1cm}

\footnotetext{\textit{$^{a}$~Sorbonne Université CNRS, MONARIS, UMR8233, F-75005 Paris, France E-mail: antoine.hacquard@sorbonne-universite.fr}}
\footnotetext{\textit{$^{b}$~Sorbonne Université, Observatoire de Paris, PSL university, CNRS, LERMA, F-75005, Paris, France}}
\footnotetext{\textit{$^{c}$~Université Paris-Saclay, CNRS, Institut des Sciences Moléculaires d'Orsay, 91405 Orsay}}
\footnotetext{\textit{$^{d}$~Univ. Lille, CNRS, UMR 8523 - PhLAM - Physique des Lasers Atomes et Molécules, F-59000 Lille, France}}
\footnotetext{\textit{$^{e}$~Physique des Interactions Ioniques et Moléculaires, CNRS, Aix Marseille Univ., 13397 Marseille, France}}

\footnotetext{\dag~Electronic Supplementary Information (ESI) available: [details of any supplementary information available should be included here]. See DOI: 00.0000/00000000.}

\footnotetext{\ddag ~Present address: Laboratory Astrophysics Group of the Max Planck Institute for Astronomy at the Friedrich Schiller University Jena,Institute of Solid State Physics, Jena, Germany }



\section{Introduction}

Non-thermal desorption is an important process in determining the astronomically observed gas phase abundances of key molecular species in the interstellar medium (ISM).
In the coldest, densest parts of the ISM, i.e. star and planet forming regions, most of the molecular matter exists as molecular ices coating interstellar dust grains. The very low temperatures (typically $\sim$ 10~K) inhibit thermal desorption, and non-thermal, energetic processes have to be invoked to interpret the observations of a significant budget of molecules in the gas phase. 
Vacuum-UV (7 - 13.6 eV) photon-induced desorption, usually refered to as UV photodesorption, is a significant non-thermal desorption pathway, especially in protoplanetary disks or at the edges of molecular clouds, where VUV radiation fields are provided either by an attenuated interstellar radiation field (ISRF), or UV radiation from the central star, penetrating the media. Additionally, secondary photons generated by cosmic rays can also be present in prestellar core, though with a different spectral profile such as a high Lyman-$\alpha$ contribution.\cite{willacy_importance_2000, hogerheijde_detection_2011, guzman_iram-30_2013, gredel_cco_1987}. 
Carbon dioxide is one of the most abundant species detected in icy grain mantles in star forming regions, along with H$_2$O and CO. In a study combining data from previous IR space observatories, the abundance of solid CO$_2$ was estimated to be 10-50\% relative to solid water ice  \citep{boogert_observations_2015}, and these numbers have since been confirmed by recent observations with the JWST.\cite{mcclure_ice_2023, dartois_spectroscopic_2024,yang__ceers_2023}  In dark clouds, CO$_2$ is found totally frozen as ice mantles\cite{bergin_gas-phase_1995} but will desorb from the grains in hot regions, for example near to massive embedded protostars\cite{boonman_gas-phase_2003}. Whereas observations and models suggest that at lower cloud densities, CO$_2$ molecules are majoritarily formed in a mixed CO$_2$:H$_2$O phase, upon reaching higher densities, a CO$_2$:CO phase begins to dominate, due to the freeze-out of gas-phase CO\cite{pontoppidan_c2d_2008}. Being a very important constituent of interstellar ices, photodesorption of CO$_2$ has attracted much attention, and motivated several experimental works these last decades in order to both better understand and quantify this phenomenon.
Among them, photodesorption processes from CO$_2$ ices have been widely studied by using microwave discharge H$_2$ flow lamps (MDHL) \cite{oberg_photodesorption_2009, bahr_photodesorption_2012, yuan_lyman-_2013, martin-domenech_uv_2015, sie_photodesorption_2019}, which are polychromatic VUV light sources  that emit photons in the 110-180~nm (6.9-11.3~eV) range, with a large contribution at 10.2~eV from the H Lyman-$\alpha$ line. The exact emission spectrum depends strongly on the conditions at which the lamps are used.\citep{chen_vacuum_2013, sie_photodesorption_2019} These studies have investigated many aspects of the photon-induced processes in CO$_2$ ices, such as for instance thickness or fluence dependencies and the photochemical evolution of the ice.  However, despite having been done in very similar conditions, the derived CO$_2$ photodesorption yields vary widely from one study to another, ranging from $1\times10^{-4}$ to $2\times10^{-2}$ ejected molecules per incident photon\cite{oberg_photodesorption_2009,sie_photodesorption_2019, martin-domenech_uv_2015,yuan_lyman-_2013}. Experiments carried out using the monochromatic and tunable VUV output of synchrotron facilities have shown that the desorption yields from CO$_2$ ices are strongly dependent on the incident photon energy, and become very important above 11.5~eV\cite{fillion_wavelength_2014}, i.e. a photon energy range which is not covered by the MDHL lamps. 
In the lower energy range, the photodesorption yields were shown not to be strongly impacted by the photon energy as solid CO$_2$ absorbs only weakly. Although the impact of the different VUV spectra of the lamps could account for the dispersion of the above-mentioned desorption yields, the effect of the photon flux on the desorption mechanisms should be considered. Although fluence effects, i.e. effects related to the total number of incident photons, were already well identified\cite{bahr_photodesorption_2012, oberg_photodesorption_2009}, the photon flux effect has not been investigated to date, despite being an important parameter that may provide information on the actual desorption mechanisms.

Different processes for the photon-induced desorption of CO$_2$ were proposed in these studies. Some invoke photochemical processes, involving dissociation and exothermic recombination of the photofragments,\citep{bahr_photodesorption_2012, sie_photodesorption_2019}  and others propose a so-called indirect desorption induced by electronic transitions (DIET) mechanism, in which energy stored in an excited molecule is transferred over some layers to a surface-located molecule, triggering its desorption.\cite{fillion_wavelength_2014, martin-domenech_uv_2015, yuan_lyman-_2013}. Insofar as some photochemical process is involved, the photodesorption yields are expected to be flux dependent, as the number of dissociation events per time and volume unit in the ice, and thus the concentration of radicals per time unit, will be strongly enhanced at higher flux conditions. In contrast, if only DIET processes are involved, no strong flux dependence is expected for the photodesorption yield, as only a single photon is involved. Both mechanisms could be simultaneously occurring, and competing, which could be at the origin of the discrepancies among the published desorption yields, as the photon flux varied greatly. Indeed, another important parameter is the photon flux used in each of the studies cited above, which varied over several orders of magnitude (typically between $10^{13}$ and $10^{15}$ photons/s/cm$^2$). Moreover, when it comes to applying the experimentally derived photodesorption yields to the interstellar medium conditions, wherein typical photon fluxes  range typically from 10$^3$ to 10$^8$ photons/s/cm$^2$,\cite{mathis_interstellar_1983, shen_cosmic_2004} that are orders of magnitude lower than in experimental conditions, and so the relationship between flux and photodesorption yield must be determined for as wide a range of flux as possible.

In this study, we have investigated photodesorption from CO$_2$ ice, using the monochromatic VUV light beam provided by the SOLEIL synchrotron (St. Aubin, France). Desorption of CO$_2$ and of other photoproducts was measured, and quantified, as a function of photon fluence using several flux conditions, varying over close to two orders of magnitude. The photochemistry of the CO$_2$ bulk ice was monitored \textit{in situ} by infrared spectroscopy in order to search for any correlation between the ice's chemical evolution and the desorption phenomena.  Here, we present the outcomes of these experimental investigations, and discuss both the involved mechanisms and the relevance of such experimental measurements to the physical conditions of the interstellar medium.

\section{Experimental method}

Photodesorption studies are performed in the SPICES (Surface Processes \& ICES) setup of the MONARIS laboratory, located in Sorbonne University (Paris, France). It is an UltraHigh Vacuum chamber with a base pressure of  around $1\times10^{-10}$ Torr. Molecular ices are grown on a gold substrate mounted on a rotatable head cooled by a closed-cycle helium cryostat. 
The temperature can be varied from 15~K to 300~K, with an absolute precision better than 0.5~K. 
Pure ices were grown at 15~K by flowing a partial pressure of gaseous CO$_2$ (Air Liquide, 99.99\% purity) through a deposition tube positioned a few millimeters in front of the cold gold substrate. The use of this method allows for the growth of relatively thick ices while limiting the increase of the base pressure in the chamber. Here, CO$_2$ ices of 50~$\pm~5$ monolayers (ML) were studied, where one monolayer corresponds to about 10$^{15}$ molecules/cm$^2$. The ice thickness was calibrated and determined by Temperature Programmed Desorption (TPD), following the protocol described in \citet{doronin_adsorption_2015}. Prior to the ice growth, the purity of the carbon dioxide was checked \textit{in situ} in the gas phase by means of mass spectrometry while keeping the temperature of the sample above 200~K. 

The SPICES setup was coupled to the beamline DESIRS (Dichro\"{i}sme et Spectroscopie par Interaction avec le Rayonnement Synchrotron)\cite{nahon_desirs_2012}  at the synchrotron facility SOLEIL (St Aubin, France). The coupling was made without any window, preventing high energy cut-off of the light beam. DESIRS is an undulator-based VUV beamline providing high photon-flux, tunable in the 5-40~eV energy range and in these experiments it was operated in the VUV range (7-14~eV). The undulator's high harmonics are suppressed by a gas filter and the radiation is sent through an 6.65~m Eagle off-plane normal-incidence monochromator, equipped for the present study with low resolution grating (200~grooves.mm$^{-1}$) allowing the adjustment of photon fluxes and spectral resolution. The photon flux can be measured using calibrated photodiodes, and varies between, at the highest, 2.2 $\times 10^{14}$~photon/s/cm$^2$ and, at the lowest, 7.3 $\times 10^{12}$~photon/s/cm$^2$, depending on the photon energy, and grating diffraction order. In fact, the possibilities to tune the flux are limited, we could only use gold grids placed on the output of the beamline, resulting in the three different flux used in our experiments. The spectral width of these photons is 0.5~eV and 15~meV when we used the zeroth or first order of diffraction of the grating of the beamline, respectively.
The irradiated part of the surface measures 1~$\pm~0.2~\text{cm}^2$. 

The neutral species desorbing into the gas phase during irradiation were monitored with a Quadrupole Mass Spectrometer (QMS, Balzers, Prisma), using electron-impact ionization at 70~eV. The mass signal is  converted into desorption yields following a protocol described in more detail below. Simultaneously, the irradiated ice was continuously monitored using Fourier Transform-Reflection Absorption Infrared Spectroscopy (RAIRS), by means of an infrared spectrometer (Bruker, vec22) providing a collimated infrared beam of diameter $\sim$20~mm forming a 72$^{\circ}$ incident angle with respect to the surface normal. Infrared spectra were all acquired with a resolution of 1~cm$^{-1}$ and 50 scans per spectrum. The probed surface by RAIRS is equivalent to the total surface of the substrate i.e. 2.25~cm$^2$. To follow the evolution of the infrared features with irradiation time, the RAIRS spectra were analyzed by fitting the observable bands, related to a given species, using a simple Gaussian shape around the maximum peak absorption of each band. The integrated absorbance was then plotted as a function of fluence as will be presented in the next sections. Finally, the Temperature Programmed Desorption (TPD) technique was used after each irradiation experiment. It consists of applying a constant heating ramp of 12 ~K/min to the sample, while measuring the desorbing species by means of mass spectrometry. TPD was employed here to estimate the chemical conversion of the ice after irradiation by monitoring the thermal desorption of CO$_2$ and of other photoproducts. 

\begin{figure}[h!]
	\begin{center}
	\includegraphics[scale=0.35]{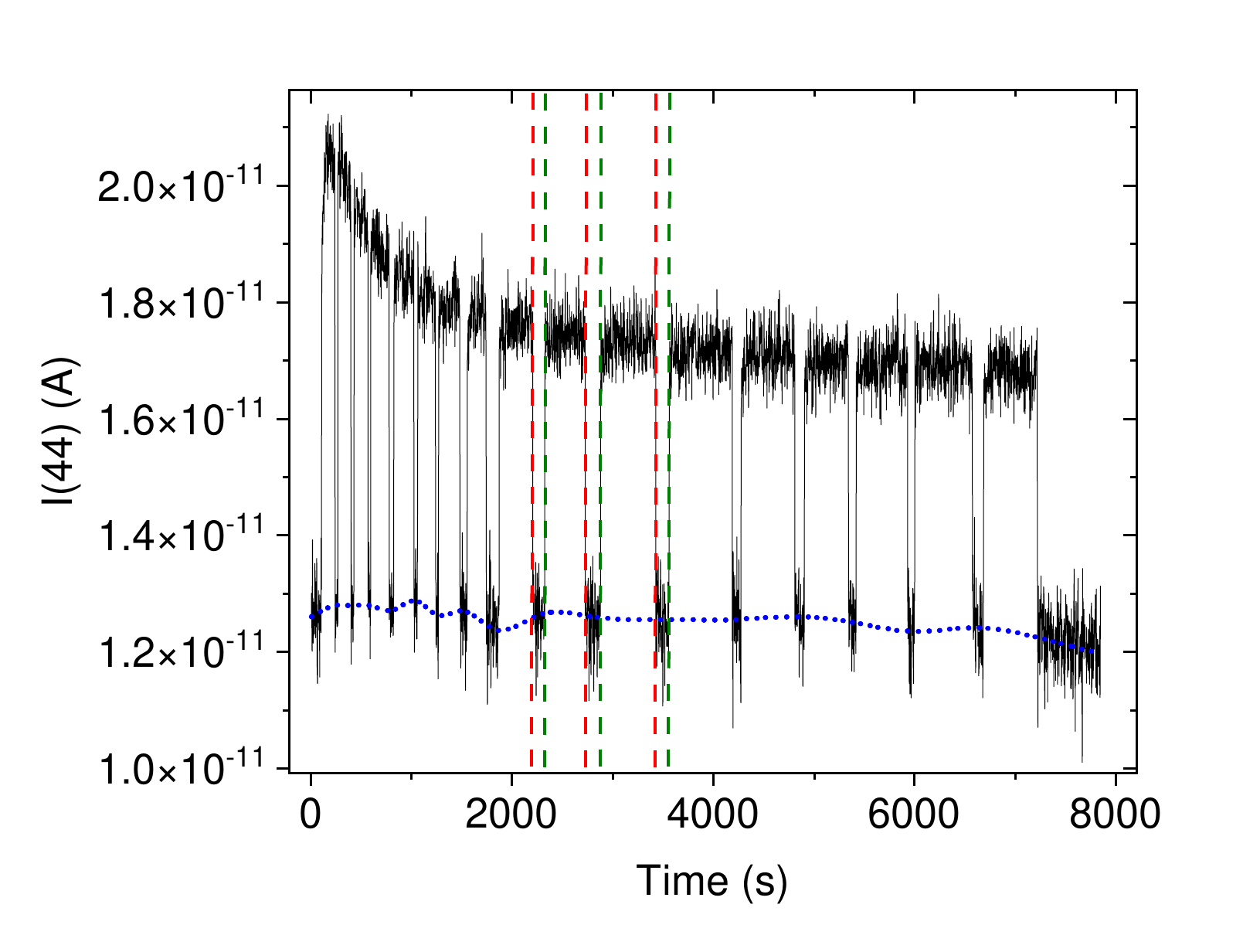}
	\caption{Raw mass signal of the channel $m=44$~u, corresponding to CO$_2$, recorded by the QMS as a function of time, during an irradiation experiment using 12~eV photons. The UV beam was regularly shut during the irradiation, in order to be able to track the evolution of the background CO$_2$ signal with time (dashed blue line). Some of the moments when the beam was shut, and then reintroduced, are shown in the figure by the vertical red and green dashed lines respectively.}
	\label{rawdata}
	\end{center}
\end{figure}

The photodesorption signal $I_X$ for an intact cation (e.g. CO$^+$ for CO, CO$_2^+$ for CO$_2$ and O$_2^+$ for O$_2$) is obtained by subtracting the continuous background signal, originating from the residual gas, from the total signal in the corresponding mass channel in order to solely isolate the signal due to the photodesorption. This is achieved by regularly shutting the UV beam during an irradiation, in order to record part of the background signal with time.
The desorption yield $\Gamma_\text{X}(E)$ at the photon energy $E$ for a species X is obtained from the mass signal of the corresponding intact cation $I_X$, measured by the QMS, and corrected by the photon flux $\Phi(E)$ following: 
\begin{equation}
\Gamma_\text{X}(E)= A_\text{X}\dfrac{I_\text{X}}{\Phi(E)}
\end{equation}
where $A_\text{X}$ is a calibration coefficient which will be discussed in the following. An example of recorded raw mass signal as a function of time is shown in Fig.~\ref{rawdata} in the case of the desorption of CO$_2$ during 12~eV irradiation. The coefficient for CO, $A_\text{CO}$ is directly deduced by recording the photodesorption signal of CO from a pure CO ice at 15~K as a function of the photon energy (in the 7-10~eV photon energy range), both in the 0th and 1st orders, then compared to the well-constrained energy-dependent photodesorption yield of pure CO taken from \citet{fayolle_co_2011}. The calibration factors for any other species $A_\text{X}$ can be obtained from $A_\text{CO}$ by correcting it by different partial electron impact ionization cross-sections at 70~eV  $\sigma(\text{X}^+/\text{X})$ and by $AF(m_\text{X})$ which is the apparatus function of the QMS, i.e. the transmission efficiency through the quadrupole which solely depends on $m_\text{X}$, the mass of the species X, as:
\begin{equation}
A_\text{X} = A_{\text{CO}} \dfrac{AF(m_\text{X})}{AF(m_\text{CO})} \dfrac{\sigma(\text{CO}^+/\text{CO})}{\sigma(\text{X}^+/\text{X})}
\end{equation}

The ionization cross-sections for CO, CO$_2$ and O$_2$ were taken from \citet{freund_measurements_1990, straub_absolute_1996}. The function $AF(m_\text{X})$ was measured for our QMS by recording gas phase mass spectra of several species, such as methanol, and comparing them to known fragmentation patterns using the NIST chemical webbook database.\cite{linstrom_nist_1997}. Finally, a similar procedure was used to correct any effect on a given mass channel due to the fragmentation of a higher mass species in the ionization chamber of the QMS, such as, for instance, a CO$^+$ signal due to the fragmentation of CO$_2$. Overall, the calibration procedure comes with systematic relative uncertainties on the desorption yields which we estimate to be about 50 \%, the main sources of error being uncertainties on the ionization cross-sections, apparatus function, background signal subtraction and photon flux determination.

\label{Experimental Set-up}

\section{Results}

\begin{figure*}[h!]
	\begin{center}
	\includegraphics[scale=0.8]{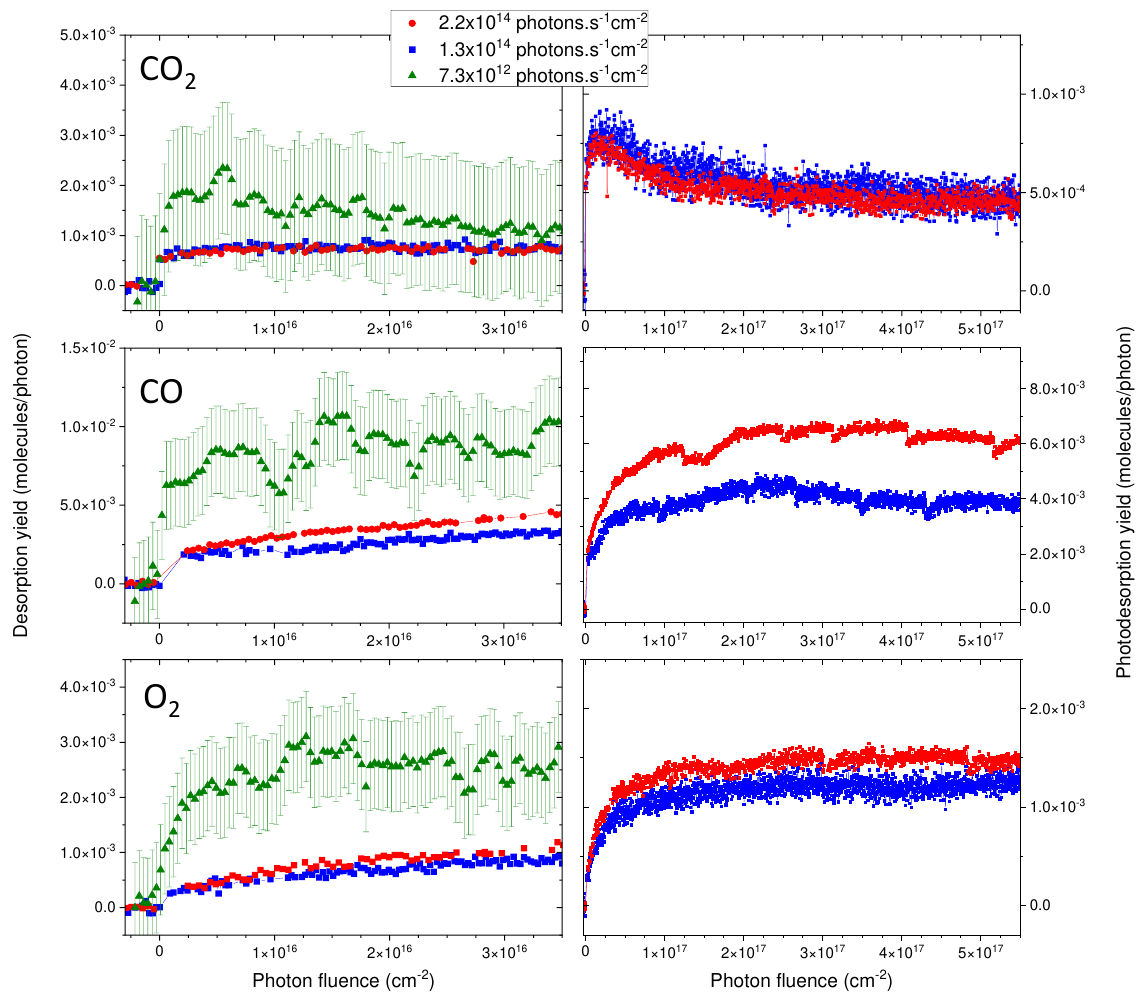}
	\caption{Photodesorption yields of  CO$_2$, CO and O$_2$ from 50~ML of pure CO$_2$ ice at 15~K, as a function of fluence during irradiation with 12~eV photons. Three different fluxes were used, namely two high flux conditions of $2.2 \times 10^{14}$~photon/s/cm$^2$ and $1.3 \times 10^{14}$~photon/s/cm$^2$ and a low flux condition of $7.3 \times 10^{12}$~photon/s/cm$^2$. The left panels show the low fluence range (up to $3.5\times10^{16}$~photons/cm$^2$) while the right panels show the high fluence regime. Breaks regularly observed on the curves, especially for the high flux yields for CO and O$_2$ are due to artifacts originating from background subtraction issues on the raw mass signals. For clarity, low flux yields are represented as mean values (over 60 points) of the desorption signals with error bars corresponding to the signal-to-noise ratio, i.e. the spread of the data points around the mean values. The low flux data are not shown on the right panel as they would appear on a very small fluence range.}
	\label{fig1}
	\end{center}
\end{figure*}

\subsection{Evolution of photon-induced desorption yields with photon fluence}

Fig. \ref{fig1} shows the photodesorption yields of CO$_2$, CO and O$_2$ from 50~ML CO$_2$ ices at 15~K, as a function of the photon fluence. These data were obtained using 12 ~eV photons, and within three flux conditions, namely a low flux regime ($7.3\times10^{12}$ photons/s/cm$^2$) and two higher flux regimes ($1.3\times10^{14}$ and $2.2\times10^{14}$ photons/s/cm$^2$). It should be noted here that, at 12~eV, all the available excited states of CO$_2$ are dissociative, and subsequent photochemistry is expected to take place. A small signal of CO$_3$ has also been identified, although very low (2.6$\times10^{-5}$~molecules/photon), however we did not detect evidence of O$_3$ photodesorption. The choice to study photodesorption at 12~eV was motivated by the fact that this photon energy falls into an energy range for which the photodesorption efficiency of CO$_2$ is at its maximum -- also corresponding to a maximum of the absorption spectrum, the associated yields dropping quickly to very low values for photon energies below 11~eV \citep{fillion_wavelength_2014}. Experiments at low photon energies, namely 8 and 10.2~eV, confirmed our previous conclusion and so will not be discussed here. At these two energies, the induced desorption due to the absorption of condensed photoproducts, for instance CO, becomes very difficult to disentangle from the desorption induced solely by CO$_2$ absorption -- very weak at these energies -- thereby complexifying the description of the evolutions with fluence.

CO$_2$ photodesorption (upper panels) builds up almost instantaneously when the irradiation begins, and exhibits at the beginning of the irradiation a value of $\sim 7\pm10^{-4}$~molecules/photon, which compares very well to previously measured yields at 12~eV \citep{fillion_wavelength_2014}. In the low fluence regime (0 -- $3.5\times10^{16}$~cm$^{-2}$), CO$_2$ photodesorption at low flux shows a slight decrease with photon fluence -- small however when compared to the uncertainties -- while the CO$_2$ photodesorption yields at higher fluxes appear almost constant with the photon fluence. The higher fluence regime (up to $6\times10^{17}$~cm$^{-2}$) shows however that the photodesorption yields of CO$_2$ at higher flux describe a small decrease toward a flat asymptotic behavior. It should be noted here that, in the case of CO$_2$ desorption, no clear flux effect could be identified. Indeed the yields for the two high flux conditions ($1.3\times10^{14}$ and $2.2\times10^{14}$~photons/s/cm$^2$) are found to be identical over the whole range of the probed fluence and cannot be compared with the low flux regime ($7.3 \times 10^{12}$~photon/s/cm$^2$) which is limited to fluences below $3.5 \times 10^{16}$~cm$^2$. Even if the lower flux condition ($7.3\times10^{12}$~photons/s/cm$^2$) seems to provide mean values for the desorption yields higher than those for the high fluxes, the very large signal-to-noise ratio does not allow us to conclude on a clear difference, since much of the lower flux yields are found equal to the higher flux yields within error bars. 

CO and O$_2$ photodesorptions both behave very similarly to one another, but are very different to the CO$_2$ case. First, the photodesorption of both CO and O$_2$ presents a continuous build-up with the photon fluence, increasing toward an asymptotic steady state at higher fluences. The dynamics seem to depend on the irradiation conditions, as fewer photons are needed to reach a steady state for low flux than for higher fluxes. Second, the yields themselves, including the values in the high fluence limit, are found to be flux dependent over the whole range of fluences studied. Interestingly, in the low fluence regime (left part of Fig.~\ref{fig1}), the lower flux value leads to higher photodesorption yields for CO and O$_2$, that reach $\sim 1\times10^{-2}$ and $2.5\times10^{-3}$~desorbed molecules per photon, respectively, for the lower flux ($7.3\times10^{12}$~photons/s/cm$^2$), compared to $6\times10^{-3}$ and $1.4\times10^{-3}$, respectively, in the high fluence regime (right part of Fig.\ref{fig1}), for the higher flux ($2.2\times10^{14}$~photons/s/cm$^2$).
When comparing now the two high flux conditions, the opposite is found, i.e. that the CO photodesorption yield is higher when the flux is increased, although this effect is not clear in the case of O$_2$.  

In order to get more information on the two observed dynamics, the fluence-dependent yields for each desorbing species ($\Gamma_{\text{CO$_2$}}$, $\Gamma_{\text{CO}}$ and $\Gamma_{\text{O$_2$}}$) at each flux were fitted with either an exponential decay in the case of CO$_2$, or an exponential growth for both CO and O$_2$, using the following single order laws:

\begin{equation}
\Gamma({\text{CO$_2$}})=[\Gamma_{\varphi_0}({\text{CO$_2$}})-\Gamma_{\infty}({\text{CO$_2$}})]e^{-\sigma_d^{\text{CO$_2$}}(\varphi-\varphi_0)} + \Gamma_{\infty}(\text{CO$_2$})
\label{eq3}
\end{equation} with $\varphi \geq \varphi_0$
and 
\begin{equation}
\Gamma_{\text{X}}=\Gamma_{\infty}({\text{X}})\left(1-e^{-\sigma_d^{\text{X}\varphi}}\right)
\label{eq4}
\end{equation} with X = CO, \text{O$_2$}

where $\varphi$ is the photon fluence, $\varphi_0$ is the photon fluence from which yields start to decrease and $\Gamma_{\infty}$ and $\sigma_d$ are fit parameters (values of $\Gamma_{\infty}$ for each species and the corresponding fits are given in the ESI). From these laws, $\sigma_d$ (obtained in the three different photon flux conditions) represents the evolution cross section of the photodesorption efficiency during the irradiation at 12~eV, a higher value of $\sigma_d$ thus indicating that fewer photons are needed to reach an asymptotic behavior. It should be emphasised here that values of $\sigma_d$ take into account the global evolution of the system, independently of the quantity and density of energy deposited in the ice. We are interested here in the global phenomena for evaluating the evolution kinetics. The fittings procedures are illustrated in Fig.~\ref{fig2} for the high flux irradiation, showing a good agreement between the experimental yields and the first order laws. The resulting values for $\sigma_d$ are shown in Table~\ref{tab:cross_sections_mass} for CO$_2$, CO and O$_2$. On the one hand, the evolution cross-sections for the CO$_2$ photodesorption appear to be very weakly dependent on the photon flux, as was already suggested from the curves in Fig.~\ref{fig1}. The obtained values vary only slightly between $9.9\times10^{-18}$~cm$^{2}$ to $6.5\times10^{-18}$~cm$^{2}$ when the photon fluxes are changed over more than an order of magnitude. On the other hand, the evolution cross-sections for both CO and O$_2$ desorption are 4 to 20 times higher (Table.~\ref{tab:cross_sections_mass}) and strongly dependent on the flux, ranging from $\sim2.5\times10^{-16}$~cm$^{2}$ in the low flux regime to an order of magnitude lower ($\sim2.4\times10^{-17}$~cm$^{2}$) in the higher flux regimes. This fact is remarkable since the more photons per second and per surface unit are used, the slower the dynamics evolves with the total number of photons. It should also be noted that the values for the evolution cross sections for the photodesorption of O$_2$ and CO are almost identical at each flux. This indicates that, for both CO and O$_2$ desorption, the mechanisms are either identical, or they involve the same physical-chemical step which control their evolution with fluence. In addition, none of the obtained values {are comparable to the cross-sections related to the evolution of CO$_2$ photodesorption, that are, depending on the flux, found between 4 to 20 times lower. This points towards a different mechanism for the CO$_2$ photodesorption, decorrelated from the desorption of its photofragments or photoproducts.

\begin{figure}[h!]
	\centering
	\includegraphics[scale=0.9]{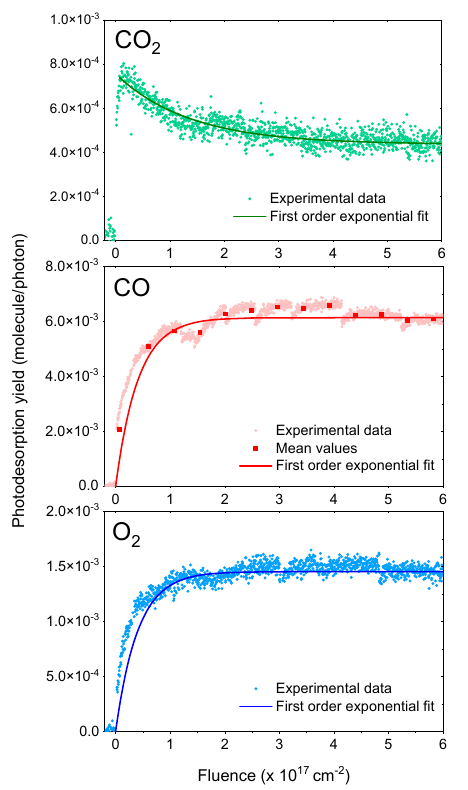}
	\caption{Photodesorption yields of CO$_2$, CO and O$_2$ from 50 ML of carbon dioxide CO$_2$ pure ice at 15 K. Kinetics are measured during irradiation with 12~eV photons and a flux of $2.2 \times 10^{14}$~photon/s/cm$^2$. The fluence-dependent yields $\Gamma(\varphi)$ are fitted using single exponential laws, either eq.\ref{eq3} for CO$_2$, or eq.\ref{eq4} for CO and O$_2$. In the case of CO, because of artefacts associated with the background substraction, mean values, obtained by averaging the yield over a short fluence range, were used.}
	\label{fig2}
\end{figure}

\begin{table*}[h!]
	\renewcommand{\arraystretch}{1.5}
	\centering
	\resizebox{0.75\textwidth}{!}{%
		\begin{tabular}{c|c|c|c}
			\hline \hline
			Photodesorption evolution cross section & \multicolumn{3}{c}{Flux (photon/s/cm$^2$)}\\
			\cline{2-4}
									for desorbing species & $7.3 \times 10^{12}$ & $1.3 \times 10^{14}$ & $2.2 \times 10^{14}$\\
			\hline
			$\sigma_d(\text{CO}_2)$  (cm$^2$) & $9.9 \pm 0.3 \times 10^{-18}$ & $6.5 \pm 0.2 \times 10^{-18}$ & $7.4 \pm 0.2 \times 10^{-18}$ \\
			\hline
			$\sigma_d(\text{CO})$ (cm$^2$) & $2.6 \pm 0.2 \times 10^{-16}$ & $2.5 \pm 0.1 \times 10^{-17}$ & $2.4 \pm 0.1 \times 10^{-17}$ \\
			\hline
			$\sigma_d(\text{O}_2)$  (cm$^2$) & $2.7 \pm 0.2 \times 10^{-16}$ & $2.1 \pm 0.1 \times 10^{-17}$ & $2.5 \pm 0.1 \times 10^{-17}$ \\
			\hline
	\end{tabular}}
	\caption{Evolution cross sections $\sigma_d$ (in cm$^2$) obtained from the first order fitting of the fluence-dependent photodesorption yields from CO$_2$ ices irradiated with 12~eV VUV photons at different photon fluxes.}
	\label{tab:cross_sections_mass}
\end{table*}

\subsection{Evolution of infrared spectra with photon fluence}

\begin{figure}[h!]
	\begin{center}
		\includegraphics[width=9cm]{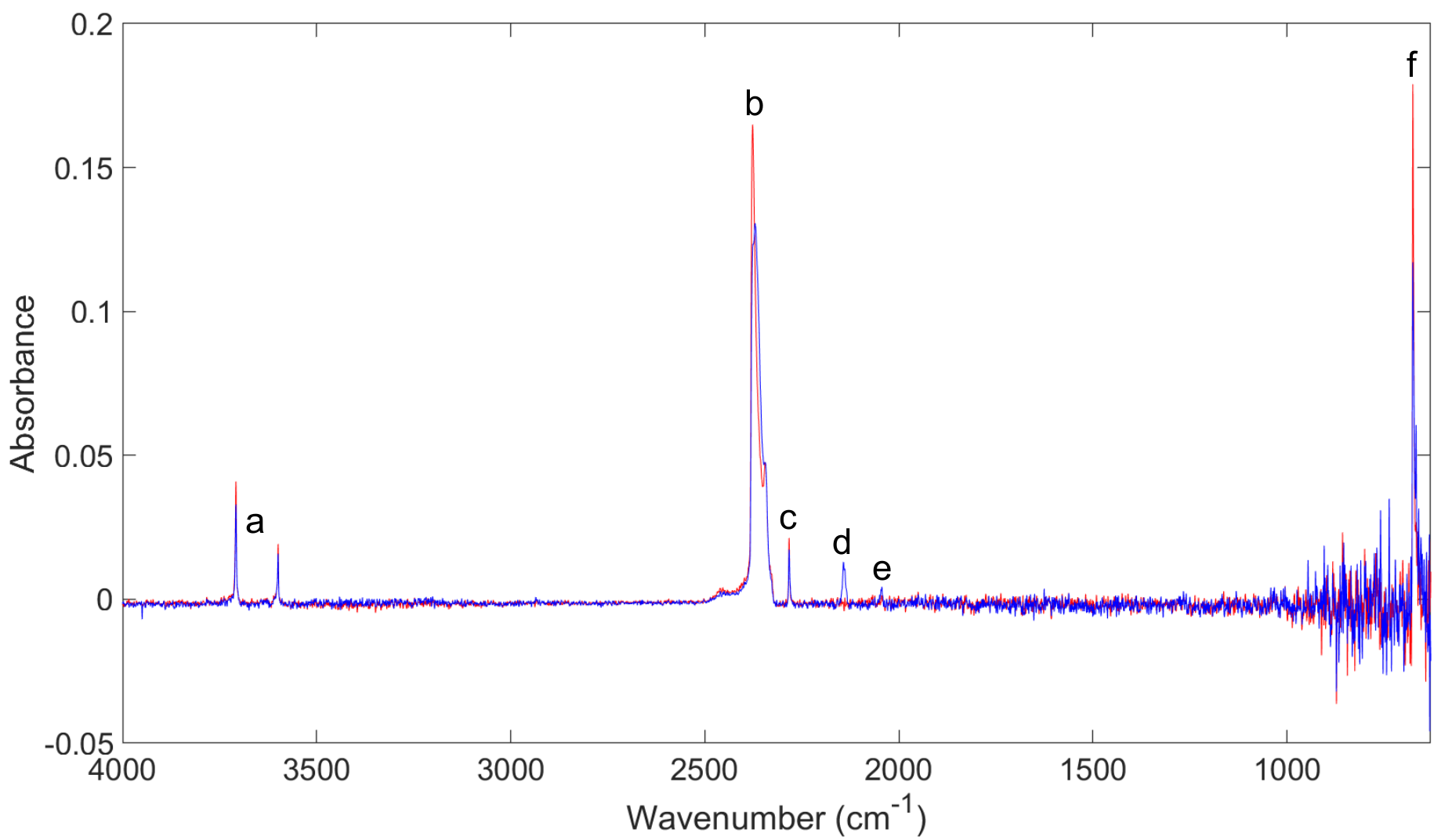}
		\caption{RAIRS spectra from 50~ML of pure CO$_2$ ice, after deposition (red spectra) and after irradiation (blue spectra) at 15~K. Ice was irradiated with photons of 12~eV and a flux of 2.2 $\times$ 10$^{14}$~photon/s/cm$^2$ and a total fluence of $1.6\times10^{18}$~cm$^{-2}$. Each observed structure has been associated with a particular vibrational mode of the solid CO$_2$ or to its photoproducts, namely (a), CO$_{2}$ combination modes ; (b), CO$_{2}$ asymetric stretching mode ; (c), $^{13}$CO$_{2}$ stretching mode ; (d), CO stretching mode ; (e), CO$_{3}$ C=O stretching mode ; (f), CO$_{2}$ bending mode.}
		\label{fig3}
	\end{center}
\end{figure}

More information on the photon-induced processes in solid CO$_2$, and their potential link to the photodesorption, can be accessed by monitoring the evolution of the photon-induced changes in the bulk ice as a function of photon fluence. Indeed, at 12~eV, the average penetration depth of the photons exceeds by far the studied ice thicknesses, and the whole ice volume should be affected by the irradiation. For this purpose, infrared spectroscopy was performed to monitor the pure CO$_2$ ices during the irradiations, as stated in Part.~\ref{Experimental Set-up}. Fig.~\ref{fig3} presents RAIRS spectra, obtained on a 50~ML CO$_2$ ice kept at 15~K, after the ice deposition, and after its irradiation using 12~eV photons with a flux of $2.2\times10^{14}$~photons/s/cm$^2$ and a total fluence of $1.6\times10^{18}$~cm$^{-2}$. Attributions of all the observed vibrational features were made using the work of \citet{gerakines_infrared_1995, escribano_crystallization_2013, baratta_infrared_1998, cooke_co_2016}. In the spectrum of the deposited ice, all the observed peaks, labeled (a), (b), (c) and (f) in Fig.\ref{fig3}, correspond to a particular vibration mode of solid CO$_2$. The CO$_2$ ${\nu}_2$  bending mode at 676~cm$^{-1}$ (f) and ${\nu}_3$ asymetric stretching mode around 2350~cm$^{-1}$ (b) are the dominant contributions to the spectrum. The ${\nu}_3$ band appears to be multi-component in our case, and will be discussed in more detailed in the following. The two bands at  3708~cm$^{-1}$ and 3600~cm$^{-1}$ (a) are associated with the combination modes ${\nu}_1+{\nu}_3$ and $2{\nu}_2+{\nu}_3$, respectively, ${\nu}_1$ being the symmetric stretching band. Finally, the band at 2283~cm$^{-1}$ (c) is related to the $\nu_3$ stretching mode of the $~^{13}$CO$_2$ isotopologue. After irradiation, the main changes are the appearance of two new vibrational bands, at 2142~cm$^{-1}$ and 2043~cm$^{-1}$, labeled (d) and (e) in Fig.~\ref{fig3}, associated with the presence of CO and CO$_3$, respectively. The band at 2043~cm$^{-1}$ related to the ${\nu_1}$ stretching mode of CO$_3$ is the only band observed for this species, other vibration modes do not emerge from the noise. A general decrease in intensity of the CO$_2$-related vibrational features is observed. This was expected as these two species were already reported to be the main photoproducts of the vacuum UV photolysis of condensed CO$_2$.\citep{gerakines_infrared_1995, martin-domenech_uv_2015, sie_photodesorption_2019} The formation of a small amount of O$_3$ was also reported in these studies, which should lead to the appearance of a vibrational band at 1043~cm$^{-1}$ related to the $\nu_3$ vibrational mode of ozone. This however could not be clearly observed in our case, most likely because of our noise level in this spectral region and of the lower fluence range which has been used here.

\begin{figure}[h!]
	\centering
	\includegraphics[width=9cm]{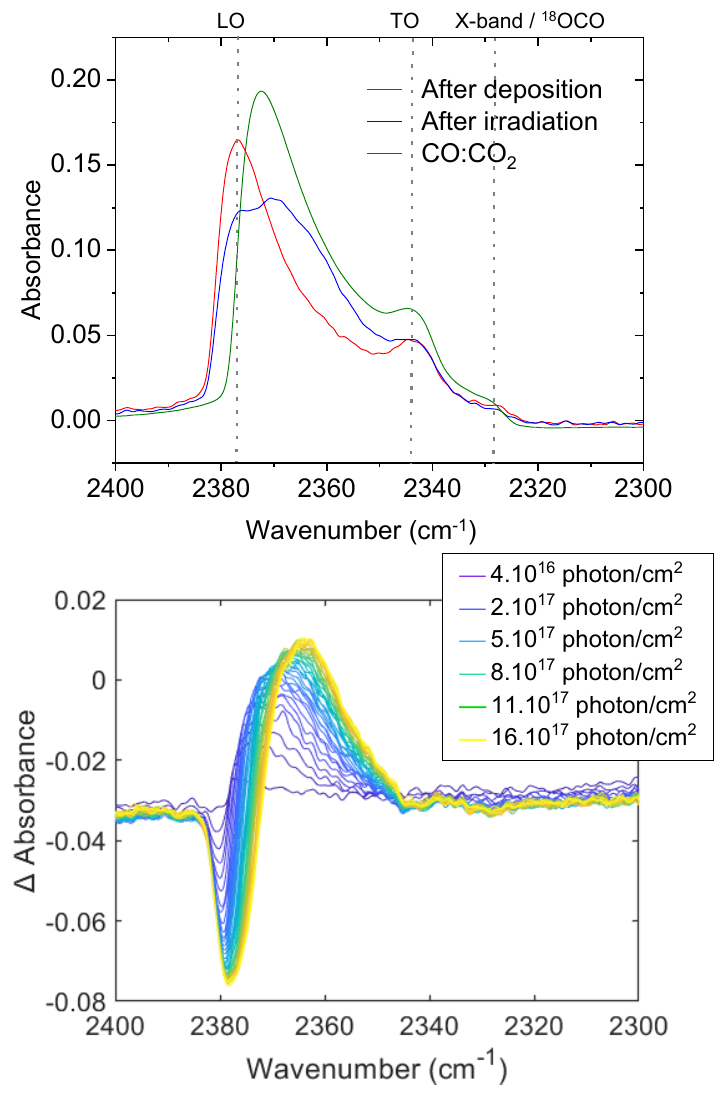}
	\caption{$\textit{Top}$~: RAIRS Spectra of the CO$_2$ stretching mode after deposition (in red) of 50~ML of CO$_2$ pure ice at 15~K and after its irradiation (in blue) with 12~eV photons and a flux of 2~$\times$~10$^{14}$~photon/s/cm$^2$, and a total fluence of $1.6\times10^{18}$~cm$^{-2}$. In green , a RAIRS spectrum of the CO$_2$ stretching mode, obtained from a mixed CO$_2$:CO 20~ML ice at 15~K, in a 4:1 proportion. The mixed CO$_2$:CO ice was prepared as a complementary experiment, with the same set-up, geometry and conditions as the pure CO$_2$ ice experiments. $\textit{Bottom}$~: Difference spectra of the evolution of the absorbance of the CO$_2$ stretching mode from 50~ML of pure ice at 15~K during VUV irradiation.}
	\label{fig4}
\end{figure}

\begin{figure}[h!]
	\centering
	
	\includegraphics[width=9cm]{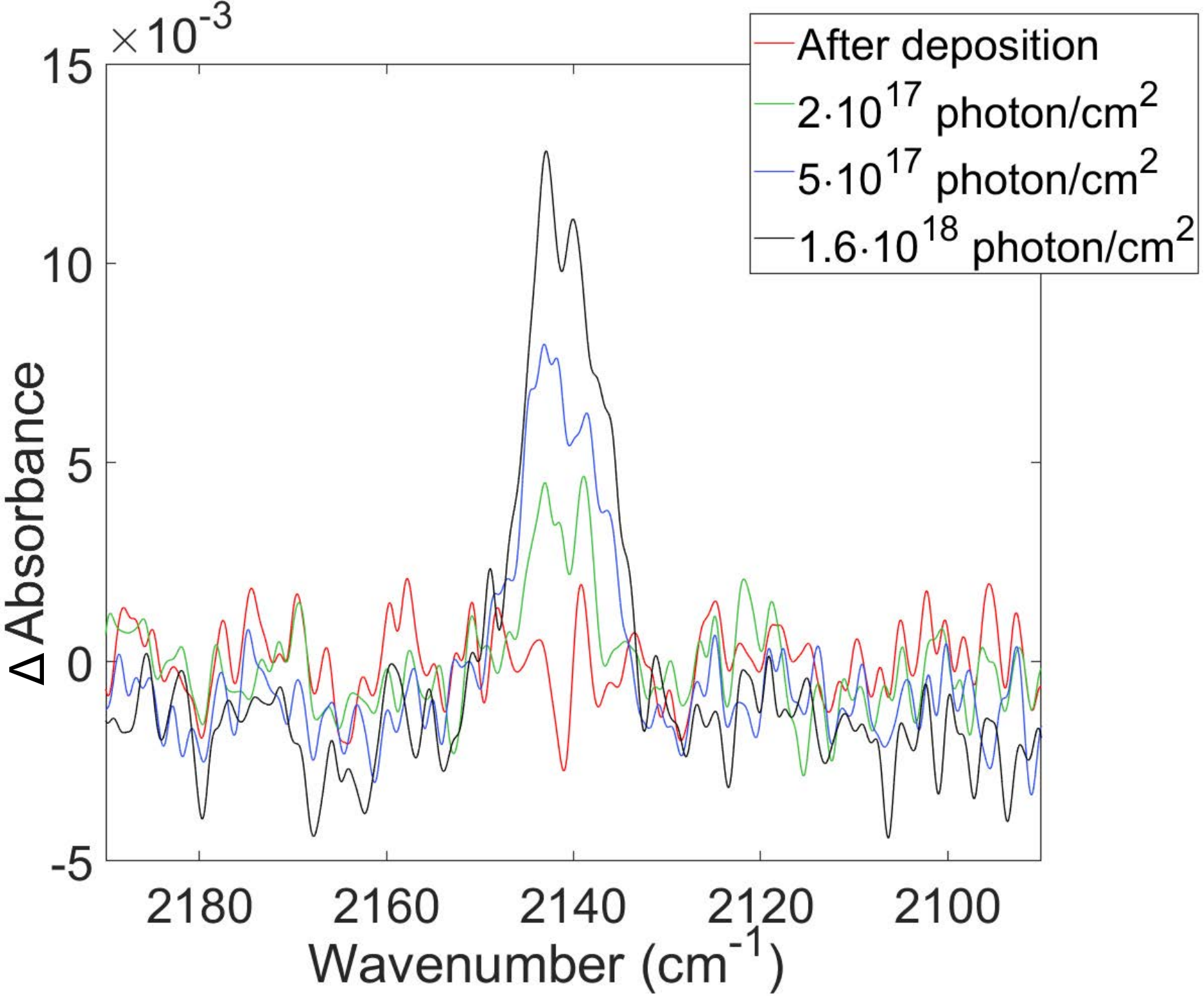}
	\caption{Difference spectra of the evolution of the absorbance of the CO stretching mode from 50~ML of pure CO$_2$ ice at 15~K during VUV irradiation with 12~eV photons and a flux of $2.2~\times~10^{14}$~photon/s/cm$^2$.}
	\label{fig5}
\end{figure}

Fig.~\ref{fig4} (top) shows the spectral range of the CO$_2$ asymmetric stretching mode after deposition of the ice, in red, and after irradiation, in blue. The lower panel shows the evolution of the difference spectra during irradiation. In our spectra, the stretching mode of deposited solid CO$_2$ consists of two main components, at 2376~cm$^{-1}$ and 2343~cm$^{-1}$. Each of these structures is associated with the excitation of a specific optical phonon mode based on the $\nu_3$ vibration mode of the isolated molecule, propagating either in the longitudinal (normal to the surface) or transverse (parallel to the surface) direction in the ice, and labeled LO and TO respectively. In our RAIRS configuration, the incident IR beam presents a strong angle with respect to the surface normal (72$^{\circ}$), and is thus able to excite both.\citep{escribano_crystallization_2013, cooke_co_2016} Finally, the shoulder at 2328~cm$^{-1}$ could be attributed to the so-called X-mode in the work of \citet{escribano_crystallization_2013}, a feature associated with the amorphous phase of solid CO$_2$. It should be noted, however, that this mode has not been observed in the RAIRS configuration in the the latter study. This feature could also be attributed to $^{18}$OCO as suggested by \citet{kataeva_infrared_2015}, and thus cannot be used as a strong marker of the amorphous character of our samples. In order to clarify this point, we have performed IR spectrum of a crystalline CO$_2$ ice grown at 50~K, which present a very different band profile than the CO$_2$ grown at 15~K (see the Electronic Supplementary Information), giving us further confidence that our sample is in an amorphous phase before irradiation. The photon irradiation mainly impacts the LO mode, the intensity of which is decreased - and bears most of the negative contributions in the difference spectra -- while the TO mode is kept almost unchanged and the X-mode only slightly modified. Another broad structure also appears at around 2360 -- 2370~cm$^{-1}$, and increases with the fluence, as can be clearly seen by the positive contribution to the difference spectra. Such structure in the CO$_2$ $\nu_3$ mode has already been observed in the study of \citet{cooke_co_2016}, and is related to CO$_2$ molecules in interaction with CO, as is shown in the upper panel of Fig.~\ref{fig4}. Indeed, the IR spectrum of an ice composed of a mixture of CO and CO$_2$ has been recorded (green line). The proportion of this CO$_2$:CO mixture has been chosen as 4:1, according to the amount of CO formed in the ice, observed in the TPD after irradiation. In our case, this contribution increases with fluence since, CO being the main photoproduct of the CO$_2$ ice processing, the ice gradually evolves from a pure CO$_2$ ice to a (predominantly) mixed CO:CO$_2$ ice, which thereby affects the global shape and position of the CO$_2$ asymmetric stretching mode. Therefore, the evolution of the $\nu_3$ band with fluence is a multiparameter process, since it reflects the total amount of CO$_2$ which is photodissociated or desorbed, together with the amount of formed CO in the ice, each contribution presenting \textit{a priori} different band strengths. Some photon-induced restructuring of the ice cannot be ruled-out either, which would induce some changes in the TO mode band strength which could compensate, in term of intensity, the global decrease of the number of CO$_2$ molecules in the ice \citep{escribano_crystallization_2013}. The fact that, after irradiation, the global band profile remains very different than the one of a CO$_2$ crystalline sample, indicates that either no crystallization occurs during the irradiation within our flux and fluence conditions, or that its signatures are negligible compared to the chemical evolution of the ice. The study of \citet{tsuge_uv-ray_2020} has shown that UV irradiation of an already crystalline CO$_2$ sample does not lead to its amorphization. They also concluded, from the study of \citet{martin-domenech_uv_2015}, that UV could further crystallize a mixture of crystalline and amorphous CO$_2$. The latter case cannot however be compared to our results on amorphous CO$_2$ as the presence of crystallites in their sample is expected to favor crystallization by nucleation process.

Due to all these overlapping contributions, the CO$_2$-related vibrational features were considered bad candidates to quantify the ice evolution with fluence, both in term of photochemistry and of photodesorption. Instead, the CO-related vibration band at 2142~cm$^{-1}$ was used in order to track the photochemical processing of the ice. We hypothesize that CO is a reliable tracer, even in the presence of other products such as O$_2$, which is invisible in the infrared. Fig.~\ref{fig5} shows the IR spectra of the deposited and of the irradiated 50~ML CO$_2$ ice at 15~K, in the spectral range of the solid CO vibration.
With ongoing fluence, the formation and accumulation of CO in the ice is observed by the continuous increase of the CO vibrational peak. The band undergoes no spectral shift nor notable changes in its width during the irradiation, which suggests that the CO molecules are in the same local environment throughout the irradiation, i.e. embedded in the CO$_2$ matrix. This was expected as, after irradiation, CO molecules account for a maximum of 20~\% of the total of the remaining species in the ice, as revealed by TPD (see the electronic supplementary information). Therefore, the intensity of the CO band can be used to trace the chemical evolution of the ice with irradiation, as we do not expect its band strength to be strongly modified with irradiation time. This evolution was monitored by plotting the integrated absorbance of the CO-related vibration peak as a function of fluence, for the pure CO$_2$ ice irradiated at 12~eV using the three previously mentioned flux conditions. Similarly to what was done for the desorption yields, each data set was then fitted with a first order single exponential law following:
\begin{equation}
a_{\text{CO}}=a_{\text{CO}}^{\infty}\left(1-e^{-\sigma_c\varphi}\right)
\label{eq5}
\end{equation}
with $a_{\text{CO}}$ being the integrated absorbance of the CO peak, $a_{\text{CO}}^{\infty}$ a fitting parameter corresponding to the asymptotic CO integrated absorbance (values are available in the ESI) and $\varphi$ the photon fluence. This law introduces the cross-section $\sigma_c$ which is defined as the evolution cross-section of the chemical modification of the CO$_2$ ice film. It should be noted that the chemical evolution which is discussed here concerns mostly the bulk of the ice: since the IR probes the whole ice thickness, the weight of the surface chemistry in the IR spectra, i.e. in the top layer at the ice-vacuum interface (estimated to correspond to $\sim$ 1~ML), can be neglected with regards to the signal related to the remaining 49~ML of the bulk ice. The associated data and fitting curves are shown in Fig.~\ref{fig6}, and the extracted values for $\sigma_c$ are presented in Table \ref{tab:cross_sections_ir}. The single exponential law well describes the behavior of the increasing band area for the two irradiations at higher flux ($1.3\times10^{14}$ and $2.2\times10^{14}$~photon/s/cm$^2$). At the lowest photon flux, no asymptotic behavior could be reached during the irradiation time, therefore we were unable to use the exponential law to fit the data. The results obtained in the high flux regime compare well with past studies from \citet{gerakines_ultraviolet_1996} who also derived a first order law for the photoproduction of CO from CO$_2$ ice irradiated with a broad-band hydrogen microwave discharge lamp. \citet{martin-domenech_uv_2015}, on the other hand, reported a decrease of the fraction of CO in the CO$_2$ ice in the high fluence regime, which they attributed to the growing importance of the recombination reaction $\text{CO} + \text{O} + \text{CO}_2 \longrightarrow 2\text{CO}_2$ competing with the CO production by direct photodissociation. This decrease was observed by \citet{martin-domenech_uv_2015} for fluences higher than $1.5\times10^{18}$~cm$^{-2}$, which is at the higher end of the fluences probed here. This might be why we do not observe this effect. For our two higher flux regimes, the kinetics for CO formation are found to be very similar, with very close evolution cross sections $\sigma_c$ of several $10^{-18}$~cm$^2$, and no clear flux effect could be identified. Additionally, similar evolution cross sections $\sigma_c$ are found in the case of CO$_3$. In addition, it is worth noting that the values we find for the chemical cross sections are found to be very similar to those reported by \citet{martin-domenech_uv_2015}, though a precise comparison should be done with caution as the photon energies were different.

\begin{figure}[h!]
	\centering
	\includegraphics[width=9cm]{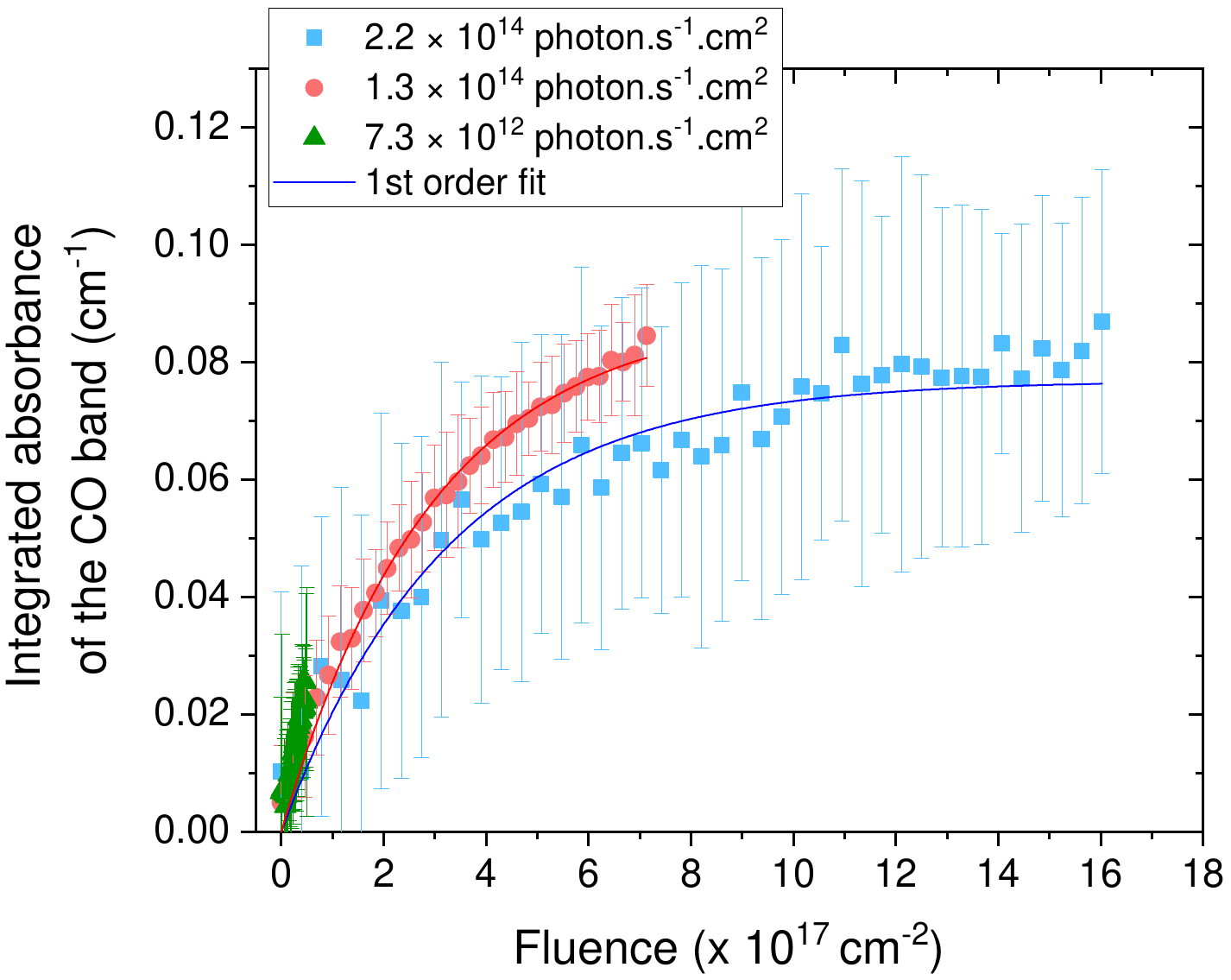}
	\caption{Evolution of the integrated absorbance of the CO vibrational band (2142~cm$^-1$) of the CO molecule as a function of the photon fluence. Irradiation was done on 50~ML of CO$_2$ pure ice samples at 15~K, with 12~eV photons for three different fluxes ($2.2 \times 10^{14}$~photon/s/cm$^2$ (blue), $1.3 \times 10^{14}$~photon/s/cm$^2$ (pink) and $7.3 \times 10^{12}$~photon/s/cm$^2$ (green). Solid lines show single exponential fitting using Eq.~\ref{eq5} for the two higher flux irradiation experiments.}
	\label{fig6}
\end{figure}

\begin{table}[h!]
	\renewcommand{\arraystretch}{1.7}
	\centering
	\resizebox{0.48\textwidth}{!}{%
		\begin{tabular}{c|c|c|c}
			\hline \hline
			Evolution cross section & \multicolumn{3}{c}{Flux (photon/cm$^2$/s)}\\
			\cline{2-4}
			for detected species & $7.3 \times 10^{12}$ & $1.3 \times 10^{14}$ & $2.2 \times 10^{14}$\\
			\hline
			$\sigma_c(\text{CO})$ (cm$^2$) & / & 3.4 $\pm$ 0.2 $\times$ 10$^{-18}$ & 3.1 $\pm$ 0.8 $\times$ 10$^{-18}$ \\
			\hline
			$a^{\infty}_{CO}$ (cm$^{-1}$) & / & 8.9 $\pm$ 0.2 $\times$ 10$^{-2}$ & 7.7 $\pm$ 0.9 $\times$ 10$^{-2}$ \\
			\hline
	\end{tabular}}
	\caption{Evolution cross sections (in cm$^2$) obtained from the integrated absorbance of the CO vibrational band of the CO photoproduct in CO$_2$ ices irradiated with different flux of VUV photon of 12~eV, using Eq.~\ref{eq5}. At low flux, cross sections cannot be extrapolated as the dynamic is too little advanced to determine an asymptotic regime.}
	\label{tab:cross_sections_ir}
	\end{table}

\section{Discussion on the photodesorption mechanisms from CO$_2$ ice}

Fluence and flux dependent experiments have highlighted two different mechanisms for the photodesorption from CO$_2$ ice. On the one hand, the photodesorption of CO$_2$ presents no clear flux-dependency, and exhibits desorption evolution cross-sections of $\sigma_d\sim6.5-10\times10^{-18}$~cm$^2$. The photochemical evolution of the CO$_2$ ice, estimated from the CO formation probed by RAIRS, exhibits the same general trend as the photodesorption of CO$_2$. Although no values could be extracted from the lower flux irradiation, chemical evolution cross sections $\sigma_c$ of the bulk ice were also found to be almost independent on the flux used in the high flux conditions, and the obtained values, i.e. $\sigma_c\sim 3-4\times10^{-18}$~cm$^2$, are close to those associated with the CO$_2$ photodesorption. 
This points toward a correlation between CO$_2$ photodesorption and the chemical evolution of the ice bulk. On the other hand, the photodesorption yields of CO and O$_2$ exhibit a strong dependence on both fluence and flux. The desorption evolution cross sections $\sigma_d$ are found to be almost identical for each species, and vary over an order of magnitude depending on the flux value, from $2.6\times10^{-16}$ to $2.4\times10^{-17}$~cm$^{2}$. In any case, these cross sections are found to be one to two orders of magnitude higher than those of the CO$_2$ desorption, and are found to be totally decorrelated from the chemical evolution of the ice bulk. In the following, we propose two mechanisms for the photodesorption from CO$_2$ ice that explain these experimental findings.

First, the photodesorption of CO$_2$ is found to be almost independent from the photon flux, which indicates a process involving a single photon absorption. The process also depends on the chemical nature of the bulk ice. This may appear surprising at first, since the photodesorbing molecules should be located at the ice surface. However, this can be explained if the process involves a single photon absorption into the bulk, followed by the transfer of the excess energy to molecules located at the ice surface. This is in fact typical of the so-called indirect desorption induced by electronic transition (DIET) process, which has previously been identified in the case of CO- or H$_2$O-containing ices.\cite{dupuy_mechanism_2021, andersson_theoretical_2011, del_fre_mechanism_2023} This type of DIET mechanism is indeed known to occur over the 3 to 5 topmost molecular layers of the ices,\cite{oberg_photodesorption_2009, munoz_caro_new_2010, bertin_uv_2012}  and its efficiency, i.e. the energy transfer to the surface, depends on the chemical nature of the layers.\cite{bertin_indirect_2013} Here, we show that the CO$_2$ photodesorption from a CO$_2$ ice is mostly due to a similar DIET process: it explains both the independence of the desorption yields from the flux, and the evolution of the efficiency with ongoing chemical modification of the bulk. For the latter, the efficiency of the DIET energy transfer evolves as the ice does, including the topmost layers, pass gradually from a pure CO$_2$ to a mixed CO:CO$_2$ environment, as revealed by the infrared monitoring. 
Our work also indicates that any photochemically-induced desorption of CO$_2$ is negligible, since it would imply flux-dependent desorption yields as will be discussed for the cases of CO and O$_2$. 

The proposed Indirect DIET mechanism for CO$_2$ photodesorption is in strong agreement with \citet{martin-domenech_uv_2015}, who concluded that the DIET process occurs in the upper layers of CO$_2$ ice during irradiation. On the other hand, \citet{sie_photodesorption_2019} claim that only CO and O$_2$ are desorbed by such an indirect DIET process. Photochemistry, even if DIET mechanism is not neglected, is believed to actively participate in the CO$_2$ desorption, in contradiction with our findings. Such a discrepancy could originate from the fact that, in the study of \citet{sie_photodesorption_2019}, the band strength of the IR stretching mode vibrational band of solid CO$_2$ was considered not to be changing with fluence, which we have shown is not the case here due to the increasing effect of the photoproduced CO in its vicinity. This, in addition to the loss of CO$_2$ due to the photochemistry, leads to inconsistent values for desorption rates measured by IR and by mass spectrometry. They corrected the photochemical effect by considering another photochemical pathway leading to the loss of C atoms. The question is open whether this process would still be needed if the fluence-induced change of the CO$_2$ bandstrength was taken into account.

The photodesorption of both CO and O$_2$ are found to originate from the same mechanism. The associated yields depend heavily on fluence and flux, and are decorrelated from the chemical evolution of the bulk ice. An hypothesis which would explain these facts is that they both originate from photon-induced surface chemistry, followed by the desorption of the photoproducts. This is supported by several points. First, as the yields depend on the photon flux, any process involving solely one photon, such as direct photodissociation of a surface CO$_2$ molecule, resulting in its fragments desorbing into the gas phase, can be ruled out. 
The increase of the photodesorption yields of CO and O$_2$ with fluence further confirms this, as it is inconsistent with a direct photodissociation of surface CO$_2$ molecules, and more indicative of an accumulation of these products on the ice surface. Second, the CO and O$_2$ desorption evolution cross-sections are found to be 10 to 100 times higher than that of the chemistry of the bulk ice. This rules-out a two step process, thereby implying (i) the formation of CO and O$_2$ at the ice surface and (ii) an indirect DIET mechanism. Indeed, as previously discussed, the DIET process involves the bulk, and it would mimic the chemical evolution of the ice, resulting in lower evolution cross sections, and is thus inconsistent with what is observed. This also reveals that the CO and O$_2$ production and desorption require many times fewer photons than the bulk chemistry, which is consistent with a surface process, where the diffusion of radicals, and thus the chemical evolution with fluence, is enhanced. To explain the CO and O$_2$ desorption mechanisms, but also the fact that the photodesorption is more efficient at low flux, we propose a surface chemical process initiated by the photodissociation of surface CO$_2$ molecules, leading to the production of hot O atoms, which can diffuse on the ice surface. It nevertheless should be noted here that our data do not provide any direct evidence of the exact photochemical pathways that are involved. In the following, we propose a scenario which fits with the above-mentioned experimental findings.

In the low flux regime, as the surface density of photon absorption events per time unit is low on the surface, the probability that the O atom encounters, and reacts with, another CO$_2$ molecule should be dominant. In this case, our hypothesis is that the resulting surface reaction is responsible for the desorption of the CO and O$_2$ via the following scheme, where ($s$) indicates a surface-located molecule, and ($g$) a molecule ejected into the gas phase:

\begin{equation}
\begin{array}{l}
\text{CO}_2(s) + h\nu \longrightarrow \text{CO}(s) + \text{O}(s)\\
 \text{O}(s) + \text{CO}_2(s) \longrightarrow  \text{CO}(g) +  \text{O}_2(g)
 \label{eq6}
 \end{array}
\end{equation}

Here, O is a produced oxygen atom which may be either translationally and/or electronically hot. The reaction $\text{O} + \text{CO}_2 \rightarrow \text{CO} + \text{O}_2$ is known to be endothermic by 340~meV\cite{pilling_mapping_2023}. The process can occur through the formation of CO$_3$, and comes with an energetic barrier of 1.53~eV for the O($^3$P) state and 481~meV for O($^1$D) state\cite{bennett_untangling_2004,grigorenko_theoretical_2020}. However, the photodissociation of CO$_2$ at 12~eV can produce oxygen atoms with excess energy up to 4.2~eV if the CO is formed in its fundamental vibrational state. At this photon energy, both hyperthermal ground state O($^3$P) with a maximum kinetic energy of 4.2~eV and excited O($^1$D) with a maximum kinetic energy of  2.1~eV, can be produced, which largely overcome the endothermicity and the barrier for the reaction. Whatever is the electronic state of the atomic oxygen, the reaction needs translationally hot oxygen atoms, as it has been reported by \citet{bennett_untangling_2004}. As the excess energy of hyperthermal O atoms are expected to be quickly dissipated, it is expected that the process involves mostly neighboring CO2 molecules, as long range diffusion of the oxygen atom would prevent the reaction.

In the higher flux regimes, the efficiency of the process decreases. This can be related to the activation of a competing surface process. Indeed, in the high flux conditions, the surface density of photon absorption events per time unit increases, and so the CO and O surface densities. These surface densities can be roughly estimated, if one considers the dissociation probability of excited CO$_2$ to be close to 1. Considering the absorption cross section $\approx 7\times10^{-18}$ cm$^2$ \citep{cruz-diaz_vacuum-uv_2014} of CO$_2$ at 12~eV, the amount of surface O and CO formed per second on the first molecular layer is of $\sim5\times10^{10}$ cm$^{-2}$ in the low flux regime, and of $\sim2\times10^{12}$ cm$^{-2}$ in the high flux regime. In the higher flux regime, the probability that an O atom, produced by a single dissociation event, meets another O atom or CO molecule originating from another dissociation event on the surface becomes higher and could compete with the O + CO$_2$ reaction that dominates at low flux regimes. The resulting O + O or O + CO reactions result in the production of O$_2$, or CO$_2$ at the surface. It has been shown by \citet{minissale_dust_2016} that neither the O + O nor O + CO reactions lead to a detectable desorption of the products (less than 5\% in their experimental conditions), which is consistent with our observations -- even if direct comparison with this study should be done with caution as the kinetic energies of the atoms are different. Therefore, this process, only efficient at high flux, should reduce the general efficiency of the desorption, as shown experimentally. The proposed competing process at high flux can be summarized as follows: 

\begin{equation}
\begin{array}{l}
\text{CO}(s) + \text{O}(s) \longrightarrow \text{CO}_2 (s) \\
\text{O}(s) + \text{O}(s) \longrightarrow \text{O}_2 (s)
\end{array}
\end{equation}   

It should be noted here that the reaction $\text{O} + \text{O}\rightarrow \text{O}_2$ was found barrierless, contrary to the reaction $\text{O} + \text{CO}\rightarrow \text{CO}_2$, and can efficiently proceed via Langmuir-Hinshelwood mechanisms at 10~K.\citep{minissale_co_2013} In our case, the molecular oxygen formation can then be induced by thermalized O atoms at 15~K, and thus be initiated by long-range diffusion of the oxygens after their formation. This process can then be operative even in the low surface densities of formed O atoms per time unit, as thermalized oxygen atoms are known to present very high lifetime at low temperature \cite{bennett_untangling_2004}. In terms of probability, it is thus more likely that the reformation of the molecular oxygen dominates gradually the surface photochemistry in the higher flux regime. The quicker chemical conversion of the surface CO$_2$ to adsorbed CO and O$_2$ competes with the low flux $\text{O} + \text{CO}_2$ reaction which leads to the desorption of CO and O$_2$.

\section{Conclusions}

In this work, we have studied the photodesorption of CO$_2$, CO and O$_2$ triggered by the irradiation of pure CO$_2$ ices with vacuum-UV photons . We have focused our study on the effects played by both the fluence and the photon flux used in the experiments, and have compared the evolution of the desorption yields with those of the chemical composition of the bulk ice as measured by infrared spectroscopy. Our work shows that CO$_2$ photodesorption proceeds mostly via a DIET mechanism, as was already hinted at in some previous experimental studies, and that the role of photochemistry is comparatively negligible, contrary to what was previously assumed. The photodesorption of CO and O$_2$ involves different mechanisms, as neither the DIET process nor direct desorption of the photofragments of surface CO$_2$ molecules can explain our findings. We propose that the desorption of these two species proceeds via surface chemistry, and involves reaction of photoproduced oxygen atom with neighboring intact CO$_2$ molecules. 

The monitoring of the photochemistry of the ice bulk using infrared spectroscopy has also revealed that the multicomponent $\nu_3$ CO$_2$ asymmetric stretching vibrational absorption band is a bad candidate to accurately follow either the chemical evolution of the ice or the matter loss due to  photodesorption. In particular, the progressive enrichment of the ice by CO molecules has a strong impact on both the spectral position, profile and intensity of the $\nu_3$ band, suggesting an important evolution of the overall band strength with the ongoing irradiation, precluding the use of the evolution of its integrated intensity with fluence as a means of quantifying the ice evolution. Some information can be drawn from it however, as our experiments have revealed that no discernible crystallization of the CO$_2$ bulk ice occurs with UV irradiation.

Our work has also demonstrated that, depending on the mechanism(s) responsible for the desorption, different photon flux conditions will lead to different measured desorption yields. While desorption induced by DIET mechanisms were shown not to depend on the photon flux, in our probed flux conditions, both the efficiency and the kinetics of the desorption with fluence are strongly impacted by the photon flux when photochemical processes are at play. First, this first shows the importance of flux-dependent approaches to disentangle between bulk or surface located processes, and, though such a method does not allow to be totally conclusive, it brings strong hints on the several photochemical pathways which may be in competition during irradiation. Second, this is a very important point to consider when it comes to extrapolating experimentally-derived photodesorption yields to cold interstellar and circumstellar environments. Indeed, typical UV photon fluxes can vary over a wide range of values in the interstellar medium. In dense molecular clouds alone, values can vary significantly: UV flux resulting from cosmic ray interactions in dense clouds\cite{shen_cosmic_2004} is, on average, approximately 10$^{3}$ to 10$^{5}$~photon/cm$^2$/s$^{-1}$, while at  the edges of the clouds, fluxes can reach approximately 10$^{7}$ to 10$^{8}$~photon/cm$^2$/s$^{-1}$ due to the attenuated interstellar radiation field (ISRF)\cite{mathis_interstellar_1983}. In any case, those interstellar fluxes are extremely low as compared to experimental UV fluxes. Most of the time, the astrophysical relevance of irradiation experiments  are discussed only in term of total fluence, compared to the integrated fluence that an icy grain can receive during the lifetime of a molecular clouds, for instance. Here, we stress that not only the fluence, but also the flux has to be taken into consideration. The flux dependence is tightly bound to the nature of the desorption mechanisms, having potentially important effects as far as photochemistry is involved.

\section*{Conflicts of interest}
There are no conflicts to declare.

\section*{Author Contributions}

\credit{AH}{1,1,1,0,1,0,0,0,0,0,0,1,1,1}
\credit{MB}{1,0,0,1,1,0,1,0,0,1,0,1,1,1}
\credit{DTD}{0,0,0,0,1,0,0,0,0,0,0,0,0,1}
\credit{RB}{0,0,0,0,1,0,0,0,0,0,0,0,0,1}
\credit{DT}{0,0,0,0,1,0,0,0,0,0,0,0,0,1}
\credit{GF}{0,0,0,0,1,0,0,0,0,0,0,0,0,1}
\credit{SD}{0,0,0,0,1,0,0,0,0,0,0,0,0,1}
\credit{JAN}{0,0,0,0,1,0,0,0,0,0,0,0,0,1}
\credit{LP}{0,0,0,0,1,0,0,0,0,0,0,0,0,1}
\credit{XM}{0,0,0,0,1,0,0,0,0,0,0,0,0,1}
\credit{JHF}{0,0,0,1,1,0,0,0,0,0,0,0,0,1}
\credit{AL}{0,0,0,0,1,0,0,0,0,0,0,0,0,1}
\credit{LM}{0,0,0,0,1,0,0,0,0,0,0,0,0,1}

\insertcreditsstatement
\section*{Acknowledgements}

	We acknowledge SOLEIL for provision of synchrotron radiation facilities under project No.~20210142 and we thank Laurent Nahon and Nelson de Oliveira for their help on the DESIRS beamline.
	This work was funded by the ANR PIXyES project, grant ANR-20-CE30-0018 of the French "Agence Nationale de la Recherche". It was supported by the DIM-ORIGINES and the DIM-ACAV+ programs of the Region Ile-de-France,  and by the Programme National “Physique et Chimie du Milieu Interstellaire" (PCMI) of CNRS/INSU with INC/INP cofunded by CEA and CNES.
	


\balance


\bibliography{refs} 

\providecommand*{\mcitethebibliography}{\thebibliography}
\csname @ifundefined\endcsname{endmcitethebibliography}
{\let\endmcitethebibliography\endthebibliography}{}
\begin{mcitethebibliography}{45}
\providecommand*{\natexlab}[1]{#1}
\providecommand*{\mciteSetBstSublistMode}[1]{}
\providecommand*{\mciteSetBstMaxWidthForm}[2]{}
\providecommand*{\mciteBstWouldAddEndPuncttrue}
  {\def\EndOfBibitem{\unskip.}}
\providecommand*{\mciteBstWouldAddEndPunctfalse}
  {\let\EndOfBibitem\relax}
\providecommand*{\mciteSetBstMidEndSepPunct}[3]{}
\providecommand*{\mciteSetBstSublistLabelBeginEnd}[3]{}
\providecommand*{\EndOfBibitem}{}
\mciteSetBstSublistMode{f}
\mciteSetBstMaxWidthForm{subitem}
{(\emph{\alph{mcitesubitemcount}})}
\mciteSetBstSublistLabelBeginEnd{\mcitemaxwidthsubitemform\space}
{\relax}{\relax}

\bibitem[Willacy and Langer(2000)]{willacy_importance_2000}
K.~Willacy and W.~D. Langer, \emph{The Astrophysical Journal}, 2000,
  \textbf{544}, 903--920\relax
\mciteBstWouldAddEndPuncttrue
\mciteSetBstMidEndSepPunct{\mcitedefaultmidpunct}
{\mcitedefaultendpunct}{\mcitedefaultseppunct}\relax
\EndOfBibitem
\bibitem[Hogerheijde \emph{et~al.}(2011)Hogerheijde, Bergin, Brinch, Cleeves,
  Fogel, Blake, Dominik, Lis, Melnick, Neufeld, Panić, Pearson, Kristensen,
  Yıldız, and Van~Dishoeck]{hogerheijde_detection_2011}
M.~R. Hogerheijde, E.~A. Bergin, C.~Brinch, L.~I. Cleeves, J.~K.~J. Fogel,
  G.~A. Blake, C.~Dominik, D.~C. Lis, G.~Melnick, D.~Neufeld, O.~Panić, J.~C.
  Pearson, L.~Kristensen, U.~A. Yıldız and E.~F. Van~Dishoeck,
  \emph{Science}, 2011, \textbf{334}, 338--340\relax
\mciteBstWouldAddEndPuncttrue
\mciteSetBstMidEndSepPunct{\mcitedefaultmidpunct}
{\mcitedefaultendpunct}{\mcitedefaultseppunct}\relax
\EndOfBibitem
\bibitem[Guzmán \emph{et~al.}(2013)Guzmán, Goicoechea, Pety, Gratier, Gerin,
  Roueff, Le~Petit, Le~Bourlot, and Faure]{guzman_iram-30_2013}
V.~V. Guzmán, J.~R. Goicoechea, J.~Pety, P.~Gratier, M.~Gerin, E.~Roueff,
  F.~Le~Petit, J.~Le~Bourlot and A.~Faure, \emph{Astronomy \& Astrophysics},
  2013, \textbf{560}, A73\relax
\mciteBstWouldAddEndPuncttrue
\mciteSetBstMidEndSepPunct{\mcitedefaultmidpunct}
{\mcitedefaultendpunct}{\mcitedefaultseppunct}\relax
\EndOfBibitem
\bibitem[Gredel \emph{et~al.}(1987)Gredel, Lepp, and Dalgarno]{gredel_cco_1987}
R.~Gredel, S.~Lepp and A.~Dalgarno, \emph{The Astrophysical Journal}, 1987,
  \textbf{323}, L137\relax
\mciteBstWouldAddEndPuncttrue
\mciteSetBstMidEndSepPunct{\mcitedefaultmidpunct}
{\mcitedefaultendpunct}{\mcitedefaultseppunct}\relax
\EndOfBibitem
\bibitem[Boogert \emph{et~al.}(2015)Boogert, Gerakines, and
  Whittet]{boogert_observations_2015}
A.~Boogert, P.~Gerakines and D.~Whittet, \emph{Annual Review of Astronomy and
  Astrophysics}, 2015, \textbf{53}, 541--581\relax
\mciteBstWouldAddEndPuncttrue
\mciteSetBstMidEndSepPunct{\mcitedefaultmidpunct}
{\mcitedefaultendpunct}{\mcitedefaultseppunct}\relax
\EndOfBibitem
\bibitem[McClure \emph{et~al.}(2023)McClure, Rocha, Pontoppidan, Crouzet, Chu,
  Dartois, Lamberts, Noble, Pendleton, Perotti, Qasim, Rachid, Smith, Sun,
  Beck, Boogert, Brown, Caselli, Charnley, Cuppen, Dickinson, Drozdovskaya,
  Egami, Erkal, Fraser, Garrod, Harsono, Ioppolo, Jiménez-Serra, Jin,
  Jørgensen, Kristensen, Lis, McCoustra, McGuire, Melnick, Öberg, Palumbo,
  Shimonishi, Sturm, Van~Dishoeck, and Linnartz]{mcclure_ice_2023}
M.~K. McClure, W.~R.~M. Rocha, K.~M. Pontoppidan, N.~Crouzet, L.~E.~U. Chu,
  E.~Dartois, T.~Lamberts, J.~A. Noble, Y.~J. Pendleton, G.~Perotti, D.~Qasim,
  M.~G. Rachid, Z.~L. Smith, F.~Sun, T.~L. Beck, A.~C.~A. Boogert, W.~A. Brown,
  P.~Caselli, S.~B. Charnley, H.~M. Cuppen, H.~Dickinson, M.~N. Drozdovskaya,
  E.~Egami, J.~Erkal, H.~Fraser, R.~T. Garrod, D.~Harsono, S.~Ioppolo,
  I.~Jiménez-Serra, M.~Jin, J.~K. Jørgensen, L.~E. Kristensen, D.~C. Lis,
  M.~R.~S. McCoustra, B.~A. McGuire, G.~J. Melnick, K.~I. Öberg, M.~E.
  Palumbo, T.~Shimonishi, J.~A. Sturm, E.~F. Van~Dishoeck and H.~Linnartz,
  \emph{Nature Astronomy}, 2023, \textbf{7}, 431--443\relax
\mciteBstWouldAddEndPuncttrue
\mciteSetBstMidEndSepPunct{\mcitedefaultmidpunct}
{\mcitedefaultendpunct}{\mcitedefaultseppunct}\relax
\EndOfBibitem
\bibitem[Dartois \emph{et~al.}(2024)Dartois, Noble, Caselli, Fraser,
  Jiménez-Serra, Maté, McClure, Melnick, Pendleton, Shimonishi, Smith, Sturm,
  Taillard, Wakelam, Boogert, Drozdovskaya, Erkal, Harsono, Herrero, Ioppolo,
  Linnartz, McGuire, Perotti, Qasim, and Rocha]{dartois_spectroscopic_2024}
E.~Dartois, J.~A. Noble, P.~Caselli, H.~J. Fraser, I.~Jiménez-Serra, B.~Maté,
  M.~K. McClure, G.~J. Melnick, Y.~J. Pendleton, T.~Shimonishi, Z.~L. Smith,
  J.~A. Sturm, A.~Taillard, V.~Wakelam, A.~C.~A. Boogert, M.~N. Drozdovskaya,
  J.~Erkal, D.~Harsono, V.~J. Herrero, S.~Ioppolo, H.~Linnartz, B.~A. McGuire,
  G.~Perotti, D.~Qasim and W.~R.~M. Rocha, \emph{Nature Astronomy}, 2024\relax
\mciteBstWouldAddEndPuncttrue
\mciteSetBstMidEndSepPunct{\mcitedefaultmidpunct}
{\mcitedefaultendpunct}{\mcitedefaultseppunct}\relax
\EndOfBibitem
\bibitem[Yang \emph{et~al.}(2023)Yang, Caputi, Papovich, Arrabal~Haro, Bagley,
  Behroozi, Bell, Bisigello, Buat, Burgarella, Cheng, Cleri, Davé, Dickinson,
  Elbaz, Ferguson, Finkelstein, Grogin, Hathi, Hirschmann, Holwerda,
  Huertas-Company, Hutchison, Iani, Kartaltepe, Kirkpatrick, Kocevski,
  Koekemoer, Kokorev, Larson, Lucas, Pérez-González, Rinaldi, Shen, Trump,
  De~La~Vega, Yung, and Zavala]{yang__ceers_2023}
G.~Yang, K.~I. Caputi, C.~Papovich, P.~Arrabal~Haro, M.~B. Bagley, P.~Behroozi,
  E.~F. Bell, L.~Bisigello, V.~Buat, D.~Burgarella, Y.~Cheng, N.~J. Cleri,
  R.~Davé, M.~Dickinson, D.~Elbaz, H.~C. Ferguson, S.~L. Finkelstein, N.~A.
  Grogin, N.~P. Hathi, M.~Hirschmann, B.~W. Holwerda, M.~Huertas-Company, T.~A.
  Hutchison, E.~Iani, J.~S. Kartaltepe, A.~Kirkpatrick, D.~D. Kocevski, A.~M.
  Koekemoer, V.~Kokorev, R.~L. Larson, R.~A. Lucas, P.~G. Pérez-González,
  P.~Rinaldi, L.~Shen, J.~R. Trump, A.~De~La~Vega, L.~Y.~A. Yung and J.~A.
  Zavala, \emph{The Astrophysical Journal Letters}, 2023, \textbf{950},
  L5\relax
\mciteBstWouldAddEndPuncttrue
\mciteSetBstMidEndSepPunct{\mcitedefaultmidpunct}
{\mcitedefaultendpunct}{\mcitedefaultseppunct}\relax
\EndOfBibitem
\bibitem[Bergin \emph{et~al.}(1995)Bergin, Langer, and
  Goldsmith]{bergin_gas-phase_1995}
E.~A. Bergin, W.~D. Langer and P.~F. Goldsmith, \emph{The Astrophysical
  Journal}, 1995, \textbf{441}, 222\relax
\mciteBstWouldAddEndPuncttrue
\mciteSetBstMidEndSepPunct{\mcitedefaultmidpunct}
{\mcitedefaultendpunct}{\mcitedefaultseppunct}\relax
\EndOfBibitem
\bibitem[Boonman \emph{et~al.}(2003)Boonman, Van~Dishoeck, Lahuis, and
  Doty]{boonman_gas-phase_2003}
A.~M.~S. Boonman, E.~F. Van~Dishoeck, F.~Lahuis and S.~D. Doty, \emph{Astronomy
  \& Astrophysics}, 2003, \textbf{399}, 1063--1072\relax
\mciteBstWouldAddEndPuncttrue
\mciteSetBstMidEndSepPunct{\mcitedefaultmidpunct}
{\mcitedefaultendpunct}{\mcitedefaultseppunct}\relax
\EndOfBibitem
\bibitem[Pontoppidan \emph{et~al.}(2008)Pontoppidan, Boogert, Fraser,
  Van~Dishoeck, Blake, Lahuis, Öberg, Evans~Ii, and
  Salyk]{pontoppidan_c2d_2008}
K.~Pontoppidan, A.~Boogert, H.~Fraser, E.~Van~Dishoeck, G.~Blake, F.~Lahuis,
  K.~Öberg, N.~Evans~Ii and C.~Salyk, \emph{The Astrophysical Journal}, 2008,
  \textbf{678}, 1005--1031\relax
\mciteBstWouldAddEndPuncttrue
\mciteSetBstMidEndSepPunct{\mcitedefaultmidpunct}
{\mcitedefaultendpunct}{\mcitedefaultseppunct}\relax
\EndOfBibitem
\bibitem[Öberg \emph{et~al.}(2009)Öberg, van Dishoeck, and
  Linnartz]{oberg_photodesorption_2009}
K.~I. Öberg, E.~F. van Dishoeck and H.~Linnartz, \emph{Astronomy \&
  Astrophysics}, 2009, \textbf{496}, 281--293\relax
\mciteBstWouldAddEndPuncttrue
\mciteSetBstMidEndSepPunct{\mcitedefaultmidpunct}
{\mcitedefaultendpunct}{\mcitedefaultseppunct}\relax
\EndOfBibitem
\bibitem[Bahr and Baragiola(2012)]{bahr_photodesorption_2012}
D.~A. Bahr and R.~A. Baragiola, \emph{The Astrophysical Journal}, 2012,
  \textbf{761}, 36\relax
\mciteBstWouldAddEndPuncttrue
\mciteSetBstMidEndSepPunct{\mcitedefaultmidpunct}
{\mcitedefaultendpunct}{\mcitedefaultseppunct}\relax
\EndOfBibitem
\bibitem[Yuan and Yates(2013)]{yuan_lyman-_2013}
C.~Yuan and J.~T. Yates, \emph{The Journal of Chemical Physics}, 2013,
  \textbf{138}, 154303\relax
\mciteBstWouldAddEndPuncttrue
\mciteSetBstMidEndSepPunct{\mcitedefaultmidpunct}
{\mcitedefaultendpunct}{\mcitedefaultseppunct}\relax
\EndOfBibitem
\bibitem[Martín-Doménech \emph{et~al.}(2015)Martín-Doménech,
  Manzano-Santamaría, Muñoz~Caro, Cruz-Díaz, Chen, Herrero, and
  Tanarro]{martin-domenech_uv_2015}
R.~Martín-Doménech, J.~Manzano-Santamaría, G.~M. Muñoz~Caro, G.~A.
  Cruz-Díaz, Y.-J. Chen, V.~J. Herrero and I.~Tanarro, \emph{Astronomy \&
  Astrophysics}, 2015, \textbf{584}, A14\relax
\mciteBstWouldAddEndPuncttrue
\mciteSetBstMidEndSepPunct{\mcitedefaultmidpunct}
{\mcitedefaultendpunct}{\mcitedefaultseppunct}\relax
\EndOfBibitem
\bibitem[Sie \emph{et~al.}(2019)Sie, Caro, Huang, Martín-Doménech, Fuente,
  and Chen]{sie_photodesorption_2019}
N.-E. Sie, G.~M.~M. Caro, Z.-H. Huang, R.~Martín-Doménech, A.~Fuente and
  Y.-J. Chen, \emph{The Astrophysical Journal}, 2019, \textbf{874}, 35\relax
\mciteBstWouldAddEndPuncttrue
\mciteSetBstMidEndSepPunct{\mcitedefaultmidpunct}
{\mcitedefaultendpunct}{\mcitedefaultseppunct}\relax
\EndOfBibitem
\bibitem[Chen \emph{et~al.}(2013)Chen, Chuang, Muñoz~Caro, Nuevo, Chu, Yih,
  Ip, and Wu]{chen_vacuum_2013}
Y.-J. Chen, K.-J. Chuang, G.~M. Muñoz~Caro, M.~Nuevo, C.-C. Chu, T.-S. Yih,
  W.-H. Ip and C.-Y.~R. Wu, \emph{The Astrophysical Journal}, 2013,
  \textbf{781}, 15\relax
\mciteBstWouldAddEndPuncttrue
\mciteSetBstMidEndSepPunct{\mcitedefaultmidpunct}
{\mcitedefaultendpunct}{\mcitedefaultseppunct}\relax
\EndOfBibitem
\bibitem[Fillion \emph{et~al.}(2014)Fillion, Fayolle, Michaut, Doronin,
  Philippe, Rakovsky, Romanzin, Champion, Öberg, Linnartz, and
  Bertin]{fillion_wavelength_2014}
J.-H. Fillion, E.~C. Fayolle, X.~Michaut, M.~Doronin, L.~Philippe, J.~Rakovsky,
  C.~Romanzin, N.~Champion, K.~I. Öberg, H.~Linnartz and M.~Bertin,
  \emph{Faraday Discussions}, 2014, \textbf{168}, 533\relax
\mciteBstWouldAddEndPuncttrue
\mciteSetBstMidEndSepPunct{\mcitedefaultmidpunct}
{\mcitedefaultendpunct}{\mcitedefaultseppunct}\relax
\EndOfBibitem
\bibitem[Mathis \emph{et~al.}(1983)Mathis, Mezger, and
  Panagia]{mathis_interstellar_1983}
J.~S. Mathis, P.~G. Mezger and N.~Panagia, \emph{Astronomy and Astrophysics},
  1983, \textbf{128}, 212--229\relax
\mciteBstWouldAddEndPuncttrue
\mciteSetBstMidEndSepPunct{\mcitedefaultmidpunct}
{\mcitedefaultendpunct}{\mcitedefaultseppunct}\relax
\EndOfBibitem
\bibitem[Shen \emph{et~al.}(2004)Shen, Greenberg, Schutte, and
  Van~Dishoeck]{shen_cosmic_2004}
C.~J. Shen, J.~M. Greenberg, W.~A. Schutte and E.~F. Van~Dishoeck,
  \emph{Astronomy \& Astrophysics}, 2004, \textbf{415}, 203--215\relax
\mciteBstWouldAddEndPuncttrue
\mciteSetBstMidEndSepPunct{\mcitedefaultmidpunct}
{\mcitedefaultendpunct}{\mcitedefaultseppunct}\relax
\EndOfBibitem
\bibitem[Doronin \emph{et~al.}(2015)Doronin, Bertin, Michaut, Philippe, and
  Fillion]{doronin_adsorption_2015}
M.~Doronin, M.~Bertin, X.~Michaut, L.~Philippe and J.-H. Fillion, \emph{The
  Journal of Chemical Physics}, 2015, \textbf{143}, 084703\relax
\mciteBstWouldAddEndPuncttrue
\mciteSetBstMidEndSepPunct{\mcitedefaultmidpunct}
{\mcitedefaultendpunct}{\mcitedefaultseppunct}\relax
\EndOfBibitem
\bibitem[Nahon \emph{et~al.}(2012)Nahon, de~Oliveira, Garcia, Gil, Pilette,
  Marcouillé, Lagarde, and Polack]{nahon_desirs_2012}
L.~Nahon, N.~de~Oliveira, G.~A. Garcia, J.-F. Gil, B.~Pilette, O.~Marcouillé,
  B.~Lagarde and F.~Polack, \emph{Journal of Synchrotron Radiation}, 2012,
  \textbf{19}, 508--520\relax
\mciteBstWouldAddEndPuncttrue
\mciteSetBstMidEndSepPunct{\mcitedefaultmidpunct}
{\mcitedefaultendpunct}{\mcitedefaultseppunct}\relax
\EndOfBibitem
\bibitem[Fayolle \emph{et~al.}(2011)Fayolle, Bertin, Romanzin, Michaut, Öberg,
  Linnartz, and Fillion]{fayolle_co_2011}
E.~C. Fayolle, M.~Bertin, C.~Romanzin, X.~Michaut, K.~I. Öberg, H.~Linnartz
  and J.-H. Fillion, \emph{The Astrophysical Journal}, 2011, \textbf{739},
  L36\relax
\mciteBstWouldAddEndPuncttrue
\mciteSetBstMidEndSepPunct{\mcitedefaultmidpunct}
{\mcitedefaultendpunct}{\mcitedefaultseppunct}\relax
\EndOfBibitem
\bibitem[Freund \emph{et~al.}(1990)Freund, Wetzel, and
  Shul]{freund_measurements_1990}
R.~S. Freund, R.~C. Wetzel and R.~J. Shul, \emph{Physical Review A}, 1990,
  \textbf{41}, 5861--5868\relax
\mciteBstWouldAddEndPuncttrue
\mciteSetBstMidEndSepPunct{\mcitedefaultmidpunct}
{\mcitedefaultendpunct}{\mcitedefaultseppunct}\relax
\EndOfBibitem
\bibitem[Straub \emph{et~al.}(1996)Straub, Renault, Lindsay, Smith, and
  Stebbings]{straub_absolute_1996}
H.~C. Straub, P.~Renault, B.~G. Lindsay, K.~A. Smith and R.~F. Stebbings,
  \emph{Physical Review A}, 1996, \textbf{54}, 2146--2153\relax
\mciteBstWouldAddEndPuncttrue
\mciteSetBstMidEndSepPunct{\mcitedefaultmidpunct}
{\mcitedefaultendpunct}{\mcitedefaultseppunct}\relax
\EndOfBibitem
\bibitem[Linstrom(1997)]{linstrom_nist_1997}
P.~Linstrom, \emph{{NIST} {Chemistry} {WebBook}, {NIST} {Standard} {Reference}
  {Database} 69}, 1997, \url{http://webbook.nist.gov/chemistry/}\relax
\mciteBstWouldAddEndPuncttrue
\mciteSetBstMidEndSepPunct{\mcitedefaultmidpunct}
{\mcitedefaultendpunct}{\mcitedefaultseppunct}\relax
\EndOfBibitem
\bibitem[Gerakines \emph{et~al.}(1995)Gerakines, Schutte, Greenberg, and van
  Dishoeck]{gerakines_infrared_1995}
P.~A. Gerakines, W.~A. Schutte, J.~M. Greenberg and E.~F. van Dishoeck,
  1995\relax
\mciteBstWouldAddEndPuncttrue
\mciteSetBstMidEndSepPunct{\mcitedefaultmidpunct}
{\mcitedefaultendpunct}{\mcitedefaultseppunct}\relax
\EndOfBibitem
\bibitem[Escribano \emph{et~al.}(2013)Escribano, Muñoz~Caro, Cruz-Diaz,
  Rodríguez-Lazcano, and Maté]{escribano_crystallization_2013}
R.~M. Escribano, G.~M. Muñoz~Caro, G.~A. Cruz-Diaz, Y.~Rodríguez-Lazcano and
  B.~Maté, \emph{Proceedings of the National Academy of Sciences}, 2013,
  \textbf{110}, 12899--12904\relax
\mciteBstWouldAddEndPuncttrue
\mciteSetBstMidEndSepPunct{\mcitedefaultmidpunct}
{\mcitedefaultendpunct}{\mcitedefaultseppunct}\relax
\EndOfBibitem
\bibitem[Baratta and Palumbo(1998)]{baratta_infrared_1998}
G.~A. Baratta and M.~E. Palumbo, \emph{Journal of the Optical Society of
  America A}, 1998, \textbf{15}, 3076\relax
\mciteBstWouldAddEndPuncttrue
\mciteSetBstMidEndSepPunct{\mcitedefaultmidpunct}
{\mcitedefaultendpunct}{\mcitedefaultseppunct}\relax
\EndOfBibitem
\bibitem[Cooke \emph{et~al.}(2016)Cooke, Fayolle, and Öberg]{cooke_co_2016}
I.~R. Cooke, E.~C. Fayolle and K.~I. Öberg, \emph{The Astrophysical Journal},
  2016, \textbf{832}, 5\relax
\mciteBstWouldAddEndPuncttrue
\mciteSetBstMidEndSepPunct{\mcitedefaultmidpunct}
{\mcitedefaultendpunct}{\mcitedefaultseppunct}\relax
\EndOfBibitem
\bibitem[Kataeva \emph{et~al.}(2015)Kataeva, Kolomiitsova, Shchepkin, and
  Asfin]{kataeva_infrared_2015}
T.~S. Kataeva, T.~D. Kolomiitsova, D.~N. Shchepkin and R.~E. Asfin,
  \emph{Chemical Physics Letters}, 2015, \textbf{641}, 117--122\relax
\mciteBstWouldAddEndPuncttrue
\mciteSetBstMidEndSepPunct{\mcitedefaultmidpunct}
{\mcitedefaultendpunct}{\mcitedefaultseppunct}\relax
\EndOfBibitem
\bibitem[Tsuge \emph{et~al.}(2020)Tsuge, Nguyen, Oba, Hama, Kouchi, and
  Watanabe]{tsuge_uv-ray_2020}
M.~Tsuge, T.~Nguyen, Y.~Oba, T.~Hama, A.~Kouchi and N.~Watanabe, \emph{Chemical
  Physics Letters}, 2020, \textbf{760}, 137999\relax
\mciteBstWouldAddEndPuncttrue
\mciteSetBstMidEndSepPunct{\mcitedefaultmidpunct}
{\mcitedefaultendpunct}{\mcitedefaultseppunct}\relax
\EndOfBibitem
\bibitem[Gerakines \emph{et~al.}(1996)Gerakines, Schutte, and
  Ehrenfreund]{gerakines_ultraviolet_1996}
P.~A. Gerakines, W.~A. Schutte and P.~Ehrenfreund, \emph{Astronomy and
  Astrophysics}, 1996, \textbf{312}, 289--305\relax
\mciteBstWouldAddEndPuncttrue
\mciteSetBstMidEndSepPunct{\mcitedefaultmidpunct}
{\mcitedefaultendpunct}{\mcitedefaultseppunct}\relax
\EndOfBibitem
\bibitem[Dupuy \emph{et~al.}(2021)Dupuy, Bertin, Féraud, Michaut,
  Marie-Jeanne, Jeseck, Philippe, Baglin, Cimino, Romanzin, and
  Fillion]{dupuy_mechanism_2021}
R.~Dupuy, M.~Bertin, G.~Féraud, X.~Michaut, P.~Marie-Jeanne, P.~Jeseck,
  L.~Philippe, V.~Baglin, R.~Cimino, C.~Romanzin and J.-H. Fillion,
  \emph{Physical Review Letters}, 2021, \textbf{126}, 156001\relax
\mciteBstWouldAddEndPuncttrue
\mciteSetBstMidEndSepPunct{\mcitedefaultmidpunct}
{\mcitedefaultendpunct}{\mcitedefaultseppunct}\relax
\EndOfBibitem
\bibitem[Andersson \emph{et~al.}(2011)Andersson, Arasa, Yabushita, Yokoyama,
  Hama, Kawasaki, Western, and Ashfold]{andersson_theoretical_2011}
S.~Andersson, C.~Arasa, A.~Yabushita, M.~Yokoyama, T.~Hama, M.~Kawasaki, C.~M.
  Western and M.~N.~R. Ashfold, \emph{Physical Chemistry Chemical Physics},
  2011, \textbf{13}, 15810\relax
\mciteBstWouldAddEndPuncttrue
\mciteSetBstMidEndSepPunct{\mcitedefaultmidpunct}
{\mcitedefaultendpunct}{\mcitedefaultseppunct}\relax
\EndOfBibitem
\bibitem[Del~Fré \emph{et~al.}(2023)Del~Fré, Santamaría, Duflot, Basalgète,
  Féraud, Bertin, Fillion, and Monnerville]{del_fre_mechanism_2023}
S.~Del~Fré, A.~R. Santamaría, D.~Duflot, R.~Basalgète, G.~Féraud,
  M.~Bertin, J.-H. Fillion and M.~Monnerville, \emph{Physical Review Letters},
  2023, \textbf{131}, 238001\relax
\mciteBstWouldAddEndPuncttrue
\mciteSetBstMidEndSepPunct{\mcitedefaultmidpunct}
{\mcitedefaultendpunct}{\mcitedefaultseppunct}\relax
\EndOfBibitem
\bibitem[Muñoz~Caro \emph{et~al.}(2010)Muñoz~Caro, Jiménez-Escobar,
  Martín-Gago, Rogero, Atienza, Puertas, Sobrado, and
  Torres-Redondo]{munoz_caro_new_2010}
G.~M. Muñoz~Caro, A.~Jiménez-Escobar, J.~A. Martín-Gago, C.~Rogero,
  C.~Atienza, S.~Puertas, J.~M. Sobrado and J.~Torres-Redondo, \emph{Astronomy
  \& Astrophysics}, 2010, \textbf{522}, A108\relax
\mciteBstWouldAddEndPuncttrue
\mciteSetBstMidEndSepPunct{\mcitedefaultmidpunct}
{\mcitedefaultendpunct}{\mcitedefaultseppunct}\relax
\EndOfBibitem
\bibitem[Bertin \emph{et~al.}(2012)Bertin, Fayolle, Romanzin, Öberg, Michaut,
  Moudens, Philippe, Jeseck, Linnartz, and Fillion]{bertin_uv_2012}
M.~Bertin, E.~C. Fayolle, C.~Romanzin, K.~I. Öberg, X.~Michaut, A.~Moudens,
  L.~Philippe, P.~Jeseck, H.~Linnartz and J.-H. Fillion, \emph{Physical
  Chemistry Chemical Physics}, 2012, \textbf{14}, 9929\relax
\mciteBstWouldAddEndPuncttrue
\mciteSetBstMidEndSepPunct{\mcitedefaultmidpunct}
{\mcitedefaultendpunct}{\mcitedefaultseppunct}\relax
\EndOfBibitem
\bibitem[Bertin \emph{et~al.}(2013)Bertin, Fayolle, Romanzin, Poderoso,
  Michaut, Philippe, Jeseck, Öberg, Linnartz, and
  Fillion]{bertin_indirect_2013}
M.~Bertin, E.~C. Fayolle, C.~Romanzin, H.~A.~M. Poderoso, X.~Michaut,
  L.~Philippe, P.~Jeseck, K.~I. Öberg, H.~Linnartz and J.-H. Fillion,
  \emph{The Astrophysical Journal}, 2013, \textbf{779}, 120\relax
\mciteBstWouldAddEndPuncttrue
\mciteSetBstMidEndSepPunct{\mcitedefaultmidpunct}
{\mcitedefaultendpunct}{\mcitedefaultseppunct}\relax
\EndOfBibitem
\bibitem[Pilling \emph{et~al.}(2023)Pilling, Rocha, Carvalho, and
  De~Abreu]{pilling_mapping_2023}
S.~Pilling, W.~R. Rocha, G.~A. Carvalho and H.~A. De~Abreu, \emph{Advances in
  Space Research}, 2023, \textbf{71}, 5466--5492\relax
\mciteBstWouldAddEndPuncttrue
\mciteSetBstMidEndSepPunct{\mcitedefaultmidpunct}
{\mcitedefaultendpunct}{\mcitedefaultseppunct}\relax
\EndOfBibitem
\bibitem[Bennett \emph{et~al.}(2004)Bennett, Jamieson, Mebel, and
  Kaiser]{bennett_untangling_2004}
C.~J. Bennett, C.~Jamieson, A.~M. Mebel and R.~I. Kaiser, \emph{Physical
  Chemistry Chemical Physics}, 2004, \textbf{6}, 735\relax
\mciteBstWouldAddEndPuncttrue
\mciteSetBstMidEndSepPunct{\mcitedefaultmidpunct}
{\mcitedefaultendpunct}{\mcitedefaultseppunct}\relax
\EndOfBibitem
\bibitem[Grigorenko \emph{et~al.}(2020)Grigorenko, Duarte, Polyakov, and
  Nemukhin]{grigorenko_theoretical_2020}
B.~L. Grigorenko, L.~Duarte, I.~V. Polyakov and A.~V. Nemukhin, \emph{Chemical
  Physics Letters}, 2020, \textbf{746}, 137303\relax
\mciteBstWouldAddEndPuncttrue
\mciteSetBstMidEndSepPunct{\mcitedefaultmidpunct}
{\mcitedefaultendpunct}{\mcitedefaultseppunct}\relax
\EndOfBibitem
\bibitem[Cruz-Diaz \emph{et~al.}(2014)Cruz-Diaz, Muñoz~Caro, Chen, and
  Yih]{cruz-diaz_vacuum-uv_2014}
G.~A. Cruz-Diaz, G.~M. Muñoz~Caro, Y.-J. Chen and T.-S. Yih, \emph{Astronomy
  \& Astrophysics}, 2014, \textbf{562}, A120\relax
\mciteBstWouldAddEndPuncttrue
\mciteSetBstMidEndSepPunct{\mcitedefaultmidpunct}
{\mcitedefaultendpunct}{\mcitedefaultseppunct}\relax
\EndOfBibitem
\bibitem[Minissale \emph{et~al.}(2016)Minissale, Dulieu, Cazaux, and
  Hocuk]{minissale_dust_2016}
M.~Minissale, F.~Dulieu, S.~Cazaux and S.~Hocuk, \emph{Astronomy \&
  Astrophysics}, 2016, \textbf{585}, A24\relax
\mciteBstWouldAddEndPuncttrue
\mciteSetBstMidEndSepPunct{\mcitedefaultmidpunct}
{\mcitedefaultendpunct}{\mcitedefaultseppunct}\relax
\EndOfBibitem
\bibitem[Minissale \emph{et~al.}(2013)Minissale, Congiu, Manicò, Pirronello,
  and Dulieu]{minissale_co_2013}
M.~Minissale, E.~Congiu, G.~Manicò, V.~Pirronello and F.~Dulieu,
  \emph{Astronomy \& Astrophysics}, 2013, \textbf{559}, A49\relax
\mciteBstWouldAddEndPuncttrue
\mciteSetBstMidEndSepPunct{\mcitedefaultmidpunct}
{\mcitedefaultendpunct}{\mcitedefaultseppunct}\relax
\EndOfBibitem
\end{mcitethebibliography}
\bibliographystyle{rsc} 

\end{document}